\documentclass[preprint,aps,12pt,notitlepage,nofootinbib,tightenlines]{revtex4}
\usepackage[utf8]{inputenc}
\usepackage{amsmath}
\usepackage{bm}
\usepackage{times}
\usepackage{braket}
\usepackage{color}
\usepackage{epsfig}
\usepackage{slashed}
\usepackage{multirow}
\usepackage{booktabs}
\usepackage{array}
\newcommand{\beq}{\begin{eqnarray}}
\newcommand{\eeq}{\end{eqnarray}}
\newcommand{\be}{\begin{equation}\begin{aligned}}
\newcommand{\ee}{\end{aligned}\end{equation}}

\newcommand{\gev}{\text{GeV}}

\begin{document}
\title{Probing $tqZ$ anomalous couplings in the trilepton signal at the HL-LHC, HE-LHC and FCC-hh}
\author{Yao-Bei Liu$^{1}$\footnote{E-mail: liuyaobei@hist.edu.cn}}
\author{Stefano Moretti$^{2}$\footnote{E-mail: s.moretti@soton.ac.uk}}
\affiliation{1. Henan Institute of Science and Technology, Xinxiang 453003, P. R. China \\  
2. School of Physics \& Astronomy, University of Southampton, Highfield, Southampton SO17 1BJ, UK }

\begin{abstract}
We investigate the prospects for discovering the Flavour Changing Neutral Current (FCNC) $tqZ$ couplings via two production processes yielding trilepton signals: top quark pair production $pp\to t\bar{t}$ with one top decaying to the $Z$ boson and one light jet and the anomalous single top plus $Z$ boson production process $pp\to tZ$. We study these channels at various successors of the Large Hadron Collider~(LHC), i.e., the approved  High-Luminosity LHC (HL-LHC) as well as the proposed
High-Energy LHC~(HE-LHC) and Future Circular Collider in hadron-hadron mode (FCC-hh). We perform a full simulation for the signals and the relevant
Standard Model (SM) backgrounds and obtain  limits on  the Branching Ratios (BRs) of $t\to qZ~(q=u,c)$, eventually yielding a trilepton final state through the decay modes $t\to b W^{+}\to b\ell^{+}\nu_{\ell}$ and $Z\to \ell^{+}\ell^{-}$. The upper limits on these FCNC BRs at 95\% Confidence Level (CL) are obtained
at the  HL-LHC with $\sqrt s=14$ TeV  and  3 ab$^{-1}$,
at the HE-LHC with $\sqrt s=27$ TeV  and 15 ab$^{-1}$ as well as
at the FCC-hh with $\sqrt s=100$ TeV and 30 ab$^{-1}$.
\end{abstract}

\maketitle

\newpage
\section{Introduction}
Being the most massive elementary particle in the Standard Model~(SM), the top quark is generally considered to be an
excellent probe for New Physics~(NP) Beyond the SM (BSM)~\cite{Tait:2000sh}. In particular, its Flavour Changing Neutral Current (FCNC) interactions are forbidden in the SM at tree-level and are strongly suppressed at loop-level by the Glashow-Iliopoulos-Maiani~(GIM) mechanism~\cite{AguilarSaavedra:2004wm,AguilarSaavedra:2009mx}.
For instance, the Branching Ratios (BRs) of $t\to qZ$ ($q=u,c$) are predicted to be at the level of $10^{-14}$ in the SM~\cite{Agashe:2013hma}, which is far
out of range of  the current Large Hadron Collider (LHC)  sensitivities.
In contrast, several NP scenarios predict the maximum values for ${\rm BR}(t\to qZ)$ $(q=u,c)$ to be at the level of
 $10^{-7}-10^{-4}$, such as the quark-singlet model~\cite{AguilarSaavedra:2002kr},
the 2-Higgs Doublet Model (2HDM) with or without flavour
conservation~\cite{Atwood:1996vj}, the Minimal Supersymmetric Standard Model~(MSSM)~\cite{Cao:2007dk}, the MSSM with $R$-parity violation~\cite{Yang:1997dk}, models with warped
extra dimensions~\cite{Agashe:2006wa} or extended mirror fermion
models~\cite{Hung:2017tts}.
Thus, searches for such FCNC processes are
very important and would be considered as a clear signal for BSM physics \cite{AguilarSaavedra:2000aj}.

Using data collected at the center-of-mass~(c.m.) energy of 13 TeV, the latest experimental limits
on the top quark FCNC  ${\rm BR}(t\to qZ)$ were established by the CMS and ATLAS collaborations by using  Run  2 data~\cite{CMS:2017twu,Aaboud:2018nyl}. The 95\% Confidence Level~(CL) upper limits are summarised in
Tab.~\ref{table:top_fcnc_results}.
As a more promising prospect, it is also worth mentioning here the scope of the
the approved  High-Luminosity LHC (HL-LHC), which is expected to reach the level of 4 to $5\times 10^{-5}$  with an integrated luminosity $L_{\rm int}$ = 3 ab$^{-1}$ at $\sqrt{s}=14$~TeV, using a full simulation of the upgraded ATLAS detector, where the three charged lepton (trilepton) final state of top quark pair events are considered, i.e., $pp\to t\bar{t}\to bW^+qZ\to b\ell \nu q \ell \ell$, where $\ell=e,\mu$~\cite{ATLAS:2019pcn}.

\begin{table} [th]
  \centering
  \caption{The current experimental upper limits on BR$(t \rightarrow qZ)$ at 95\% CL.}\label{table:top_fcnc_results}\vspace*{0.25cm} %
  \renewcommand{\arraystretch}{1.5}
  \begin{tabular}{c |c | c| c } \hline\hline
  \textbf{Detector} & BR$(t \rightarrow u Z)$ & BR$(t \rightarrow c Z)$ & Ref. \\ \hline
CMS, 13 TeV, 35.9 fb$^{-1}$ & $2.4 \times 10^{-4}$ & $4.5 \times 10^{-4}$&~\cite{CMS:2017twu}  \\
\hline
  ATLAS, 13 Tev,  36.1 fb$^{-1}$& $1.7 \times 10^{-4}$ & $2.4 \times 10^{-4}$ &~\cite{Aaboud:2018nyl}  \\ \hline
  \end{tabular}
\end{table}

At present, therefore, there is no experimental evidence of such top quark FCNC anomalous couplings.
One can however improve these limits, or indeed achieve  discovery,  at  future higher  luminosity and/or higher energy hadron colliders~\cite{Mandrik:2018yhe}, such as the aforementioned HL-LHC and/or
the proposed High-Energy LHC~(HE-LHC), with 27 TeV of c.m. energy and 15 ab$^{-1}$ of integrated luminosity~\cite{Benedikt:2018ofy} as well as the Future Circular Collider in hadron-hadron mode~(FCC-hh), with 100 TeV of c.m. energy and 30 ab$^{-1}$ of integrated luminosity ~\cite{Arkani-Hamed:2015vfh}.

\begin{figure}[!t]\vspace{0.5cm}
\begin{center}
\centerline{\epsfxsize=12cm \epsffile{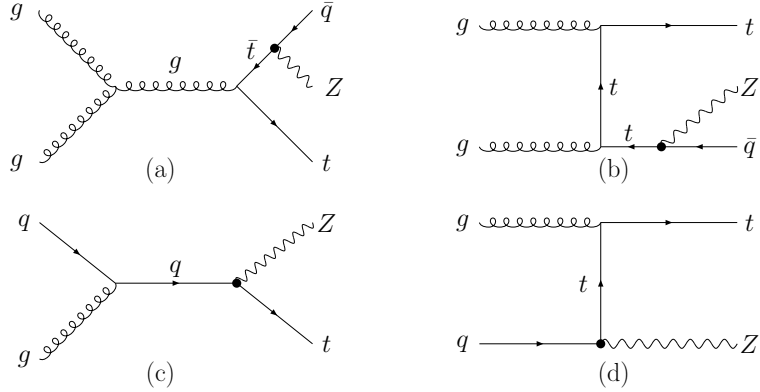}}
\vspace{-10cm}
\caption{Representative Feynman diagrams for $t\bar{t}\to t  \bar{q} Z$ production and decay (a--b) and $tZ$ associated production (c--d), both of which proceed  via FCNC $tZq$ anomalous couplings ($q=u,c$).}
\label{fey}
\end{center}
\end{figure}

The aim of this work is to investigate the limits on the discussed  $tqZ$ anomalous couplings that can be placed at these future hadron colliders using a trilepton signature. In fact, in addition to the latter being generated via $t\bar{t}$ production followed by the  FCNC $t\to qZ$ decay mode~(hereafter, $t\bar{t}$-FCNC), also  single top quark production in association with a $Z$ boson~(hereafter, $tZ$-FCNC)  leads to a trilepton signature~\cite{Agram:2013koa,Shen:2018mlj}, albeit with no hard jets stemming from the hard scattering, as shown in Fig.~\ref{fey}. Following the approach of Ref.~\cite{Liu:2020kxt} for the case of FCNC $tqh$ ($q=u,c$) anomalous couplings (wherein $h$ is the SM-like Higgs boson discovered at the LHC in 2012), also in this paper,  the search for FCNC $tqZ$ anomalous couplings is performed by combining the above two processes in the trilepton final state, where both the $W^\pm$ boson from the top quark and the $Z$ boson decay into either electrons or muons.
Thus, we consider two different trilepton signal selections, one where at least two  jets with at least one $b$-tag are required (corresponding to the
$t\bar{t}$-FCNC channel) and  the other where exactly one $b$-tagged jet is required (corresponding  to the $tZ$-FCNC channel).
Realistic detector effects are included in both signal and background processes, so that
the emerging results can be compared to  experimental predictions.

This paper is arranged as follows. In Sec.~II, the cross sections of the the two signal processes are calculated at the discussed  hadron colliders. Then Sec.~III includes estimates for the signal and background event rates alongside  95\% CL limits on the advocated trilepton signals.  Finally, we   summarise our main results and conclude in Sec.~IV.


\section{Production and decay processes with top quark FCNC interactions}
In this section, we describe the structure of the $tqZ$ interactions and quantify the cross sections of the production and decay processes of interest here involving these.
\subsection{The FCNC $tqZ$ anomalous couplings}
In the search for  FCNC $tqZ$ anomalous interactions,
the top quark FCNC coupling is explored in a model-independent way by considering the most general effective Lagrangian approach~\cite{AguilarSaavedra:2008zc}.
The  Lagrangian involving FCNC $tqZ$  interactions
can be written as~\cite{AguilarSaavedra:2008zc}
\begin{eqnarray}
\label{lag}
\mathcal{L}_{\rm eff} &=&\sum_{q=u,c} [\frac{g}{4c_{W}m_{Z}}\kappa_{tqZ} \bar{q}\sigma^{\mu\nu}(\kappa_{L}P_{L}+\kappa_{R}P_{R})tZ_{\mu\nu}  \nonumber \\
&+& \frac{g}{2c_{W}}\lambda_{tqZ} \bar{q}\gamma^{\mu}(\lambda_{L}P_{L}+\lambda_{R}P_{R})tZ_{\mu}] + h.c.,
\end{eqnarray}
where $c_{W}=\cos\theta_{W}$ and $\theta_{W}$ is the Weinberg
angle, $P_{L,R}$ are the left- and right-handed chirality projector operators, $\kappa_{tqZ}$ and $\lambda_{tqZ}$ are effective couplings for the corresponding vertices. The Electro-Weak (EW) interaction is parameterised by the coupling constant
$g$ and the mixing angle $\theta_{W}$. The complex chiral parameters $\kappa_{L,R}$ and $\lambda_{L,R}$
are normalised as $|\kappa_{L}|^2 + |\kappa_{R}|^2=|\lambda_{L}|^2 + |\lambda_{R}|^2=1$.

 The partial widths for the FCNC decays, wherein we separate the contributions of the two tensor structures entering the above equation,  are given by
 \beq
\Gamma(t\to qZ)~(\sigma^{\mu\nu})&=&\frac{\alpha}{128s_{W}^{2}c_{W}^{2}}|\kappa_{tqZ}|^2\frac{m_{t}^{3}}{m_{Z}^{2}}\left[1-\frac{m_{Z}^{2}}{m_{t}^{2}}\right]^{2}\left[2+\frac{m_{Z}^{2}}{m_{t}^{2}}\right], \nonumber \\
\Gamma(t\to qZ)~(\gamma^{\mu})&=&\frac{\alpha}{32s_{W}^{2}c_{W}^{2}}|\lambda_{tqZ}|^2\frac{m_{t}^{3}}{m_{Z}^{2}}\left[1-\frac{m_{Z}^{2}}{m_{t}^{2}}\right]^{2}\left[1+2\frac{m_{Z}^{2}}{m_{t}^{2}}\right]. \eeq
After neglecting all the light quark masses and assuming the dominant top decay partial width to be that of $t \to bW$~\cite{Li:1990qf}
\beq
\Gamma(t\to bW^{+})=\frac{\alpha}{16s_{W}^{2}}|V_{tb}|^2\frac{m_{t}^{3}}{m_{W}^{2}}\left[1-3\frac{m_{W}^{4}}{m_{t}^{4}}+2\frac{m_{W}^{6}}{m_{t}^{6}}\right], \eeq
then the BR($t \to qZ$) can be approximated  by~\cite{AguilarSaavedra:2004wm}
 \beq
{\rm BR}(t \to qZ)~(\sigma^{\mu\nu})&=&0.172|\kappa_{tqZ}|^2,\\ \nonumber
{\rm BR}(t \to qZ)~(\gamma^{\mu})&=&0.471|\lambda_{tqZ}|^2.
\eeq
Here, the Next-to-Leading Order   (NLO) QCD corrections to the top quark decay via model-independent FCNC couplings are also included and the $k$-factor is taken as 1.02~\cite{Zhang:2008yn,Drobnak:2010wh}.

\subsection{Cross sections}
For the simulations of the ensuing collider phenomenology, we first use the FeynRules package~\cite{feynrules} to generate the Universal
FeynRules Output~(UFO) files~\cite{Degrande:2011ua}. The LO cross sections are obtained by using MadGraph5-aMC$@$NLO~\cite{Alwall:2014hca} with NNPDF23L01 Parton Distribution Functions (PDFs)~\cite{Ball:2014uwa} taking the renormalisation and factorisation scales to be $\mu_R=\mu_F=\mu_0/2=(m_t + m_Z)/2$.
The numerical values of the input parameters are taken as follows~\cite{pdg}:
\begin{align}
 m_t&=173.1{\rm ~GeV},\quad m_Z=91.1876{\rm ~GeV}, \quad m_W=80.379{\rm ~GeV},\\ \nonumber
\alpha_{s}(m_Z)&=0.1181, \quad G_F=1.16637\times 10^{-5}\ {\rm GeV^{-2}}.
\end{align}

\begin{figure}[!t]
\begin{center}
\vspace{-0.5cm}
\centerline{\epsfxsize=7cm\epsffile{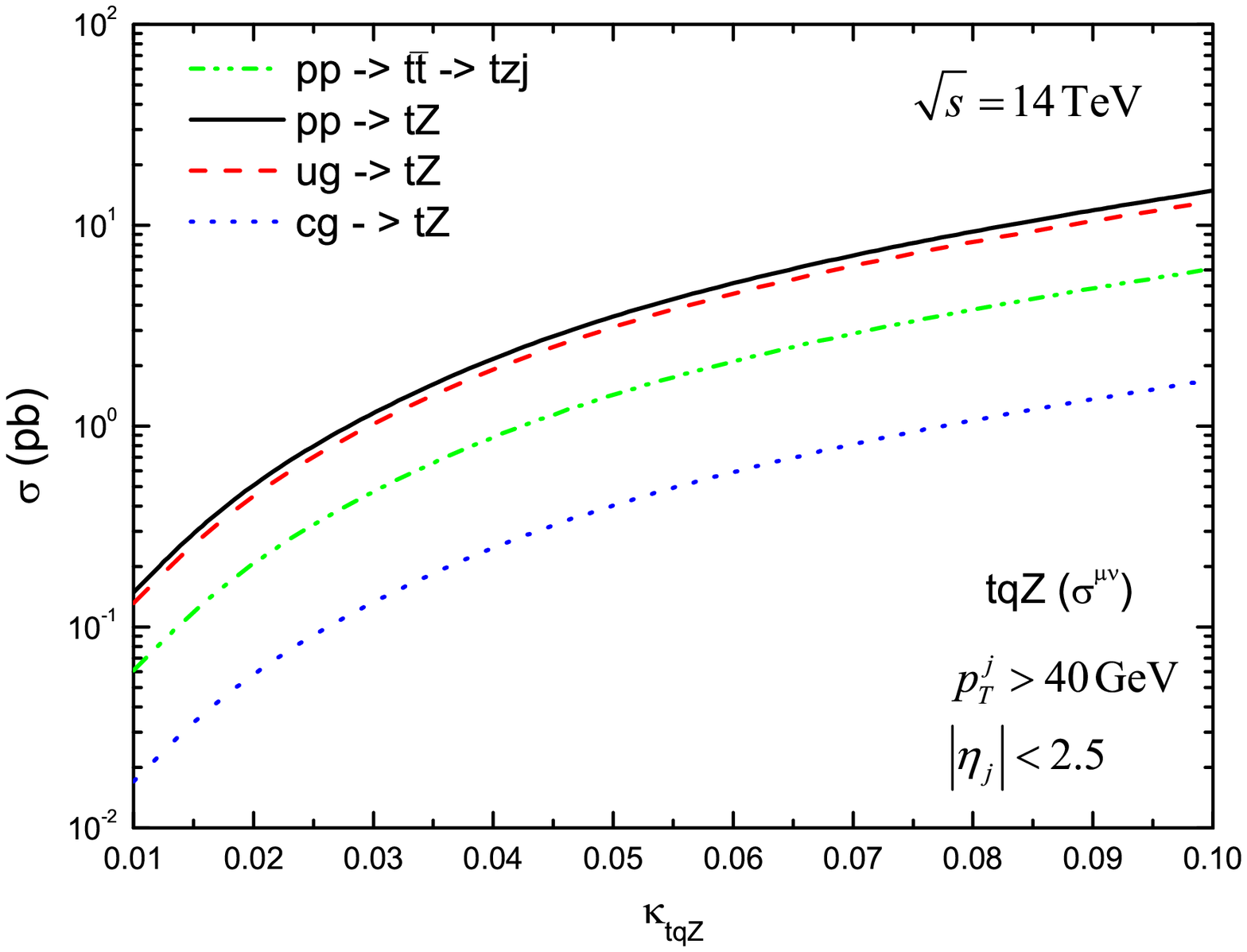}
\hspace{-1.5cm}\epsfxsize=7cm\epsffile{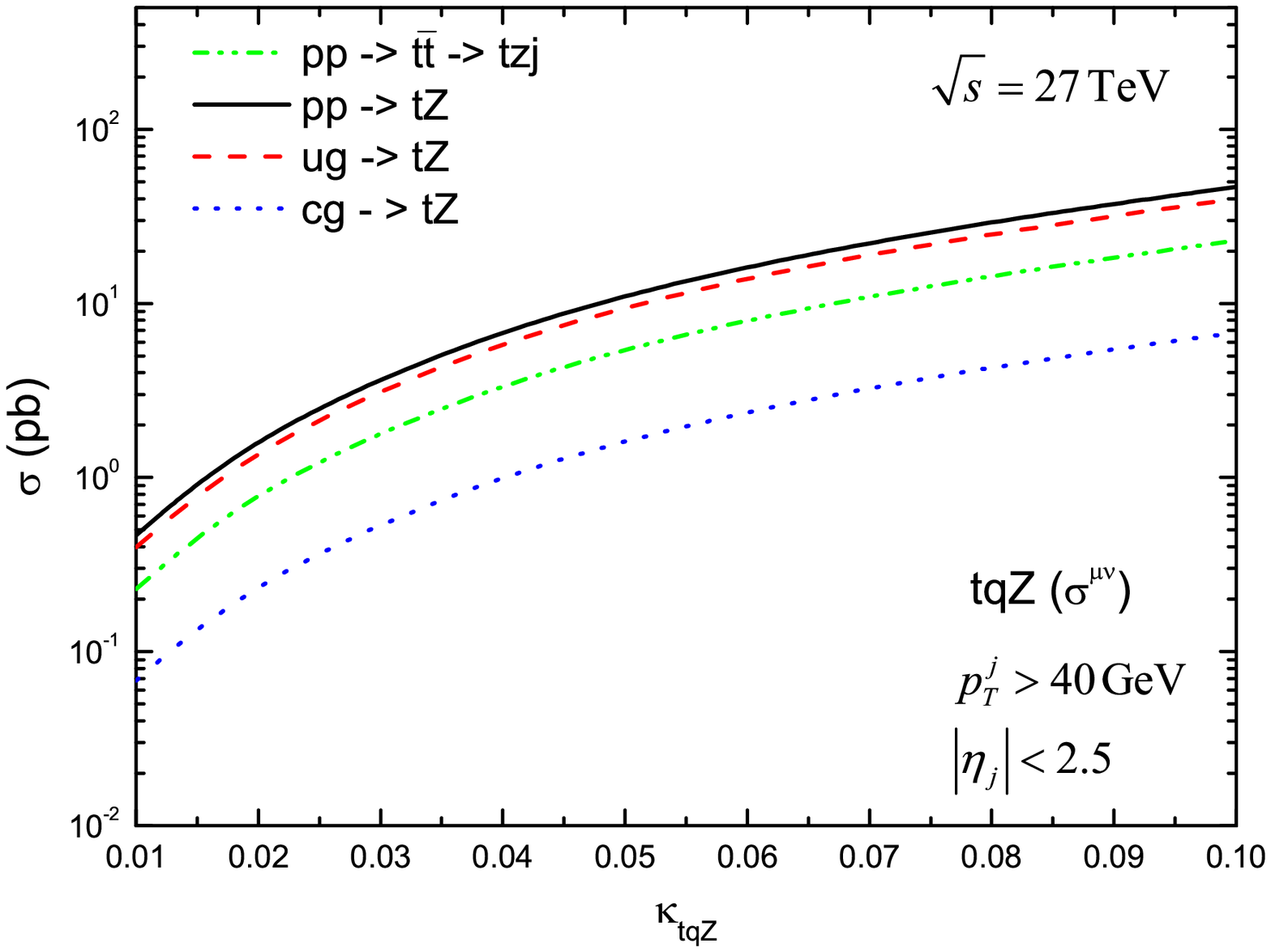}
\hspace{-1.5cm}\epsfxsize=7cm\epsffile{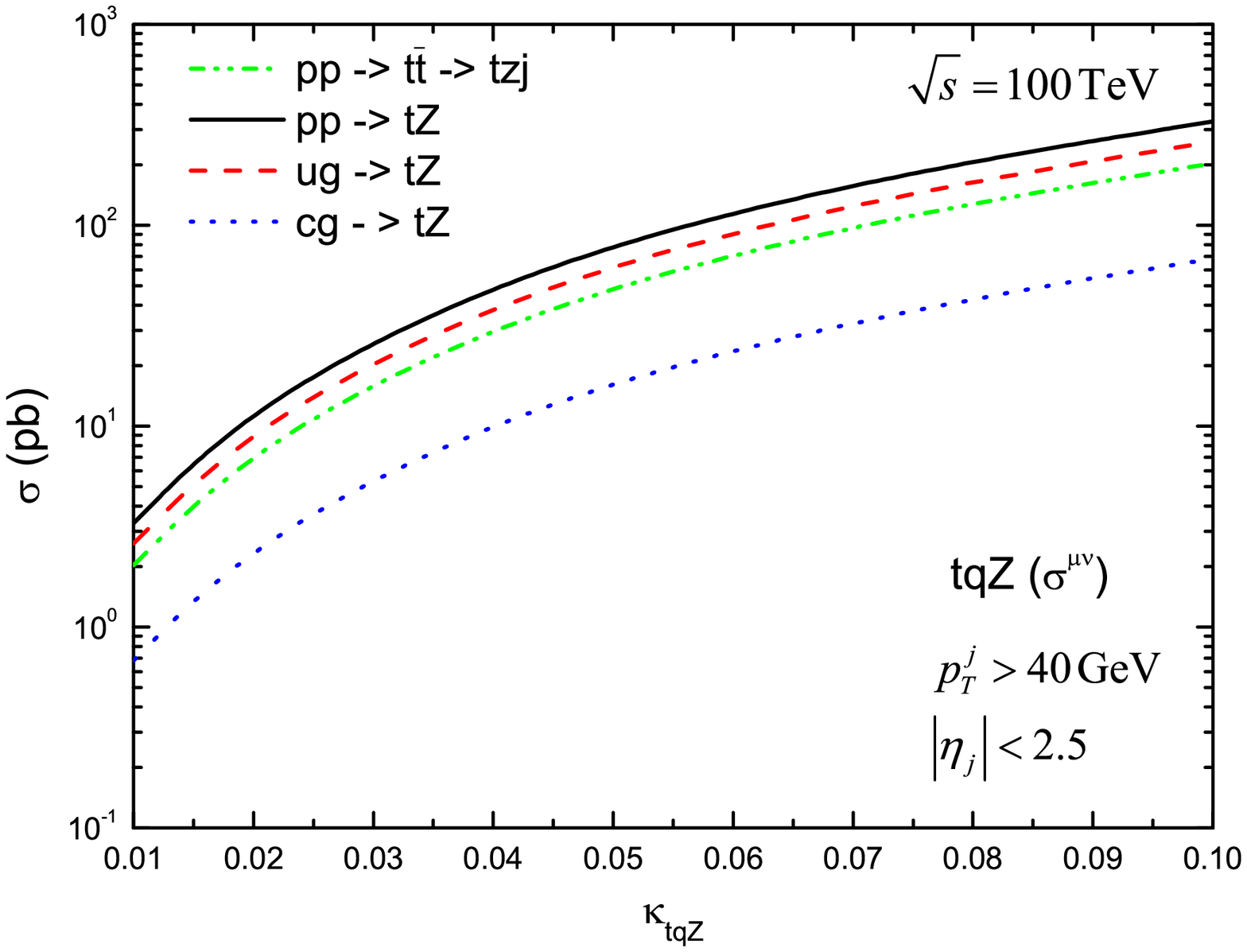}}
\centerline{\epsfxsize=7cm\epsffile{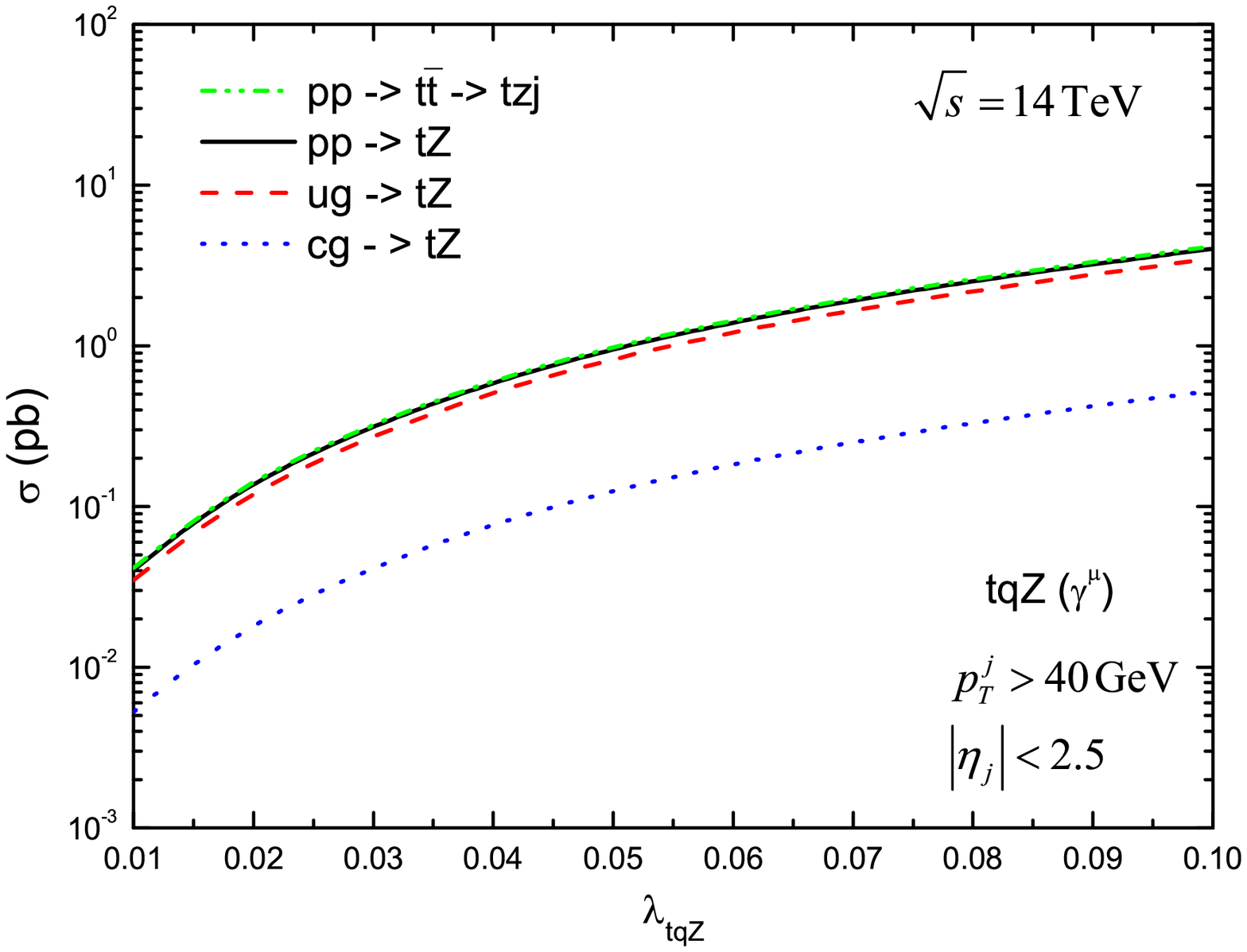}
\hspace{-1.5cm}\epsfxsize=7cm\epsffile{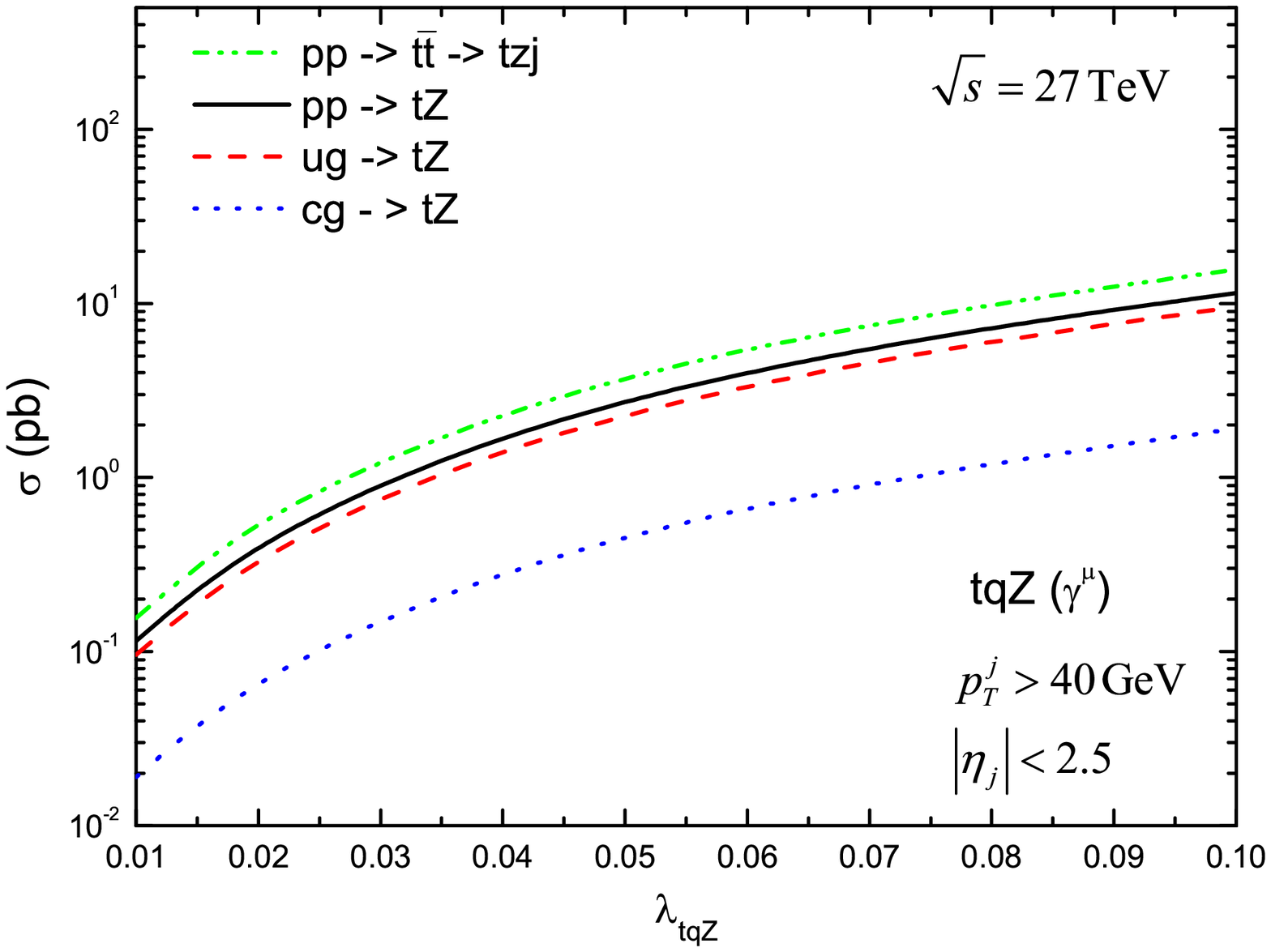}
\hspace{-1.5cm}\epsfxsize=7cm\epsffile{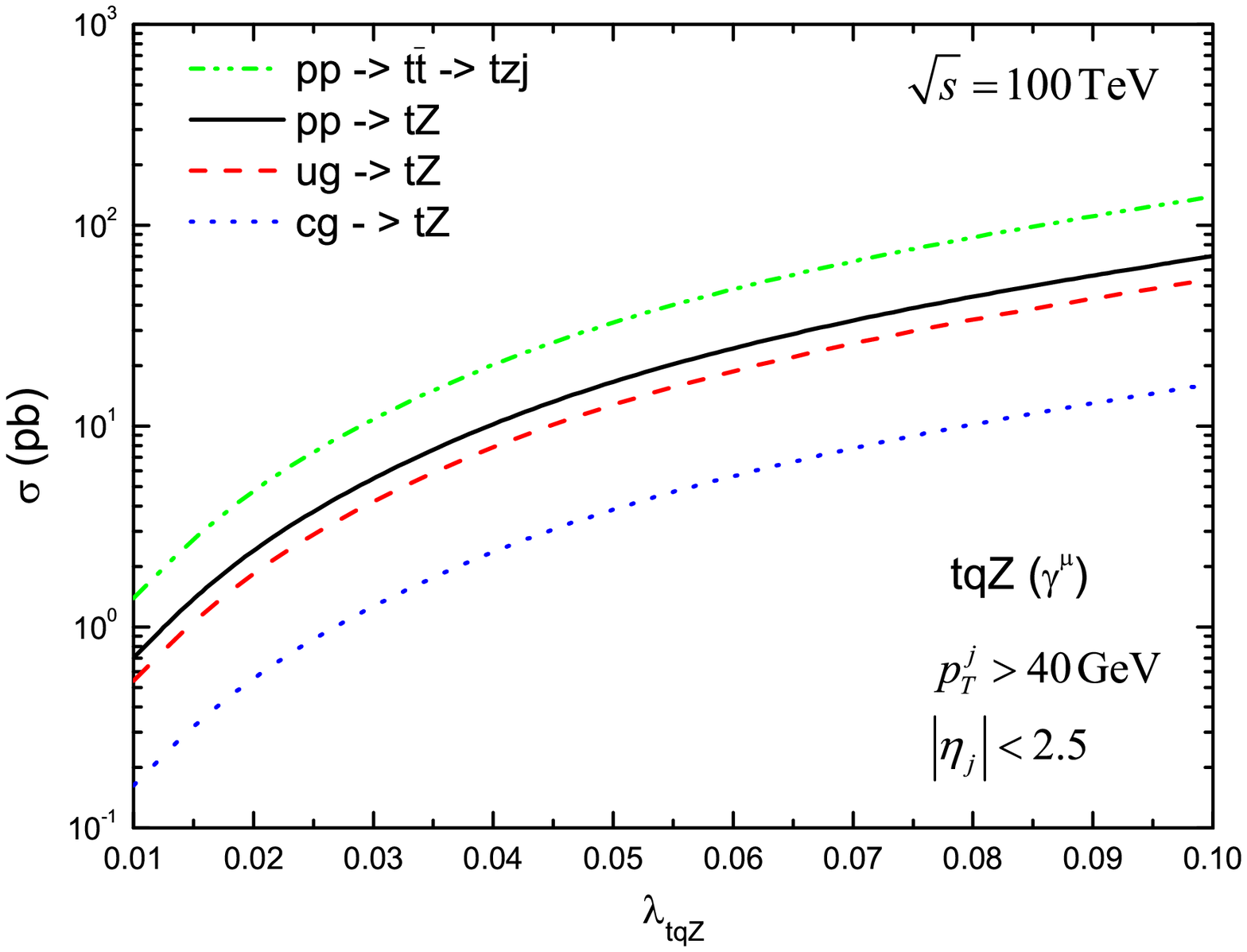}}
\caption{The dependence of the cross section $\sigma$ on the FCNC coupling parameters $\kappa_{tqZ}$~(upper)
and $\lambda_{tqZ}$~(lower) at the HL-LHC~(left), HE-LHC~(middle) and FCC-hh~(right) with the basic cuts: $p_T^j>40$~GeV and $|\eta_j|< 2.5$. Notice that the charge conjugated  processes are also included in the calculation.}
\label{cross}
\end{center}
\end{figure}

In Fig.~\ref{cross}, we show the total cross sections $\sigma$ in pb versus the two kinds of coupling parameters, $\kappa_{tqZ}$ and $\lambda_{tqZ}$, at LO. One can see that the dipole $\sigma^{\mu\nu}$ terms lead to larger cross sections with the same coupling values. For the  two kinds of couplings, the cross sections of $\bar{u}g\to \bar{t}Z$ are overwhelmed by $ug\to tZ$ due to the difference between the $u$-quark and $\bar{u}$-quark PDF of the proton. Thus, if we consider the leptonic top decay modes, more leptons will be observed than anti-leptons for a given c.m. energy and integrated luminosity. Due to the similarly small PDFs of the $c$-quark and $\bar{c}$-quark, the cross section of $\bar{c}g\to \bar{t}Z$ is essentially the same as that of $cg\to tZ$  and they are much smaller than the cross section of $ug\to tZ$ for the same values of the coupling parameter. This implies that the sensitivity to the FCNC coupling parameter $\kappa_{tuZ}~(\lambda_{tuZ})$ will be better than $\kappa_{tcZ}~(\lambda_{tcZ})$.

\section{Simulation and analysis}
In this section, we describe the numerical treatment of our signal and background events.
\subsection{The signal and background analysis}
The signal is  produced through the processes (herein, all charge conjugated channels are included)
\beq\label{signal}
pp&\to &t(\to b W^{+}\to b\ell^{+}\nu)Z(\to \ell^{+}\ell^{-}),\\
pp&\to &t(\to b W^{+}\to b \ell^{+}\nu)\bar t(\to \bar q Z(\to \ell^{+}\ell^{-})),
\eeq
where $\ell= e, \mu$ and $q=u,c$, the latter   eventually generating a jet $j$.

 As intimated, the final state for the signal is characterised by three leptons~(electrons and/or muons), one $b$-tagged jet plus missing transverse energy from the escaping undetected neutrino in the $tZ$-FCNC case. In the final state from the $t\bar{t}$-FCNC process, there is an additional jet arising from the hadronisation of the quark $q$. Furthermore, notice that
the interference between the $tZ$-FCNC (with an additional $q$ emission) and $t\bar{t}$-FCNC processes can be neglected~\cite{Barros:2019wxe}.

  The main backgrounds which yield identical final states to the signal ones are $W^\pm Z$ production in association with jets, $t\bar{t}V$ ($V=W^\pm , Z$) and the irreducible $tZj$ process, where $j$ denotes a non-$b$-quark jet. Besides, in the top pair production case (where the top quark pairs decay semi-leptonically), a third lepton can come from a semi-leptonic $B$-hadron decay inside the $b$-jet. Here, we do not consider multijet backgrounds where jets can fake  electrons, since they are generally negligible in multilepton analyses~\cite{Khachatryan:2014ewa}.  Other processes, such as the  $t\bar{t}h$, tri-boson events or $W^\pm$ + jets are not included in the analysis due to the very small cross sections after applying the cuts.

The signal and background samples are generated at LO by interfacing   MadGraph5-aMC$@$NLO to the
the Monte Carlo (MC) event generator
 Pythia 8.20~\cite{Sjostrand:2014zea}  for the parton showering.
All produced jets are forced to be clustered using FASTJET 3.2~\cite{Cacciari:2011ma} assuming the anti-$k_{t}$ algorithm with a cone radius of $R=0.4$~\cite{Cacciari:2008gp}. All event samples are fed into the Delphes 3.4.2 package~\cite{deFavereau:2013fsa} with the default HL-LHC, HE-LHC and FCC-hh detector cards. Finally, the event analysis is performed by using MadAnalysis5~\cite{Conte:2012fm}.
 To include inclusive QCD contributions, we generate
the hard scatterings of signal and backgrounds with up to
one additional jet in the final state, followed by matrix
element and parton shower merging with the MLM
matching scheme~\cite{Frederix:2012ps}. Furthermore,
 we renormalise the LO cross sections for the signals to the corresponding higher order QCD results of Refs.~\cite{Li:2011ek,Zhang:2010dr,Degrande:2014tta}.
For the SM backgrounds, we generated LO samples renormalised to the NLO or next-NLO (NNLO) order cross sections, where available, taken from Refs.~\cite{Lazopoulos:2008de,Kardos:2011na,Czakon:2013goa,Mangano:2016jyj,Campanario:2010hp,Campbell:2012dh,Frederix:2017wme,Frixione:2015zaa,Pagani:2020mov,Azzi:2019yne}.
For instance, the LO cross section for the $W^{\pm}Z$ + jets background (one of the most relevant ones overall) is renormalised to the NLO one through a $k$-factor of 1.3~\cite{Campanario:2010hp} at 14 TeV LHC and, as an estimate, we assume the same correction factor at the HE-LHC and FCC-hh. The LO cross section for the $t\bar{t}$ process is renormalised to the NNLO one by a $k$-factor of 1.6~\cite{Azzi:2019yne} at the HL-LHC as well as HE-LHC  and 1.43~\cite{Mangano:2016jyj} at the FCC-hh.

In order to identify objects, we impose the following basic or generation (parton level) cuts for the signals and SM backgrounds:
 \be
p_{T}^{\ell}>25~\gev,\quad
p_{T}^{j/b}>30~\gev,\quad
 |\eta_{i}|<2.5,\quad
 \Delta R_{ij}>0.4 ~~(i,j = \ell, b, j),
 \ee
where $j$ and $b$ denote light-flavour jets and a $b$-tagged jet, respectively. Here, $\Delta R=\sqrt{\Delta\Phi^{2}+\Delta\eta^{2}}$ is the separation in the rapidity-azimuth plane. Next, we discuss the events selection by focusing on two cases:
the $pp \to t\bar{t}\to tZj$~(henceforth referred to as `Case A') process and the $pp \to tZ$~(henceforth referred to as `Case B') process, respectively. As mentioned, the main difference is whether there is a light jet in the final state. We first discuss the selection cuts for Case A and then for Case B.

\subsection{The  selection cuts for Case A}

\begin{figure}[htb]
\begin{center}
\centerline{\hspace{1.5cm}\epsfxsize=9.5cm\epsffile{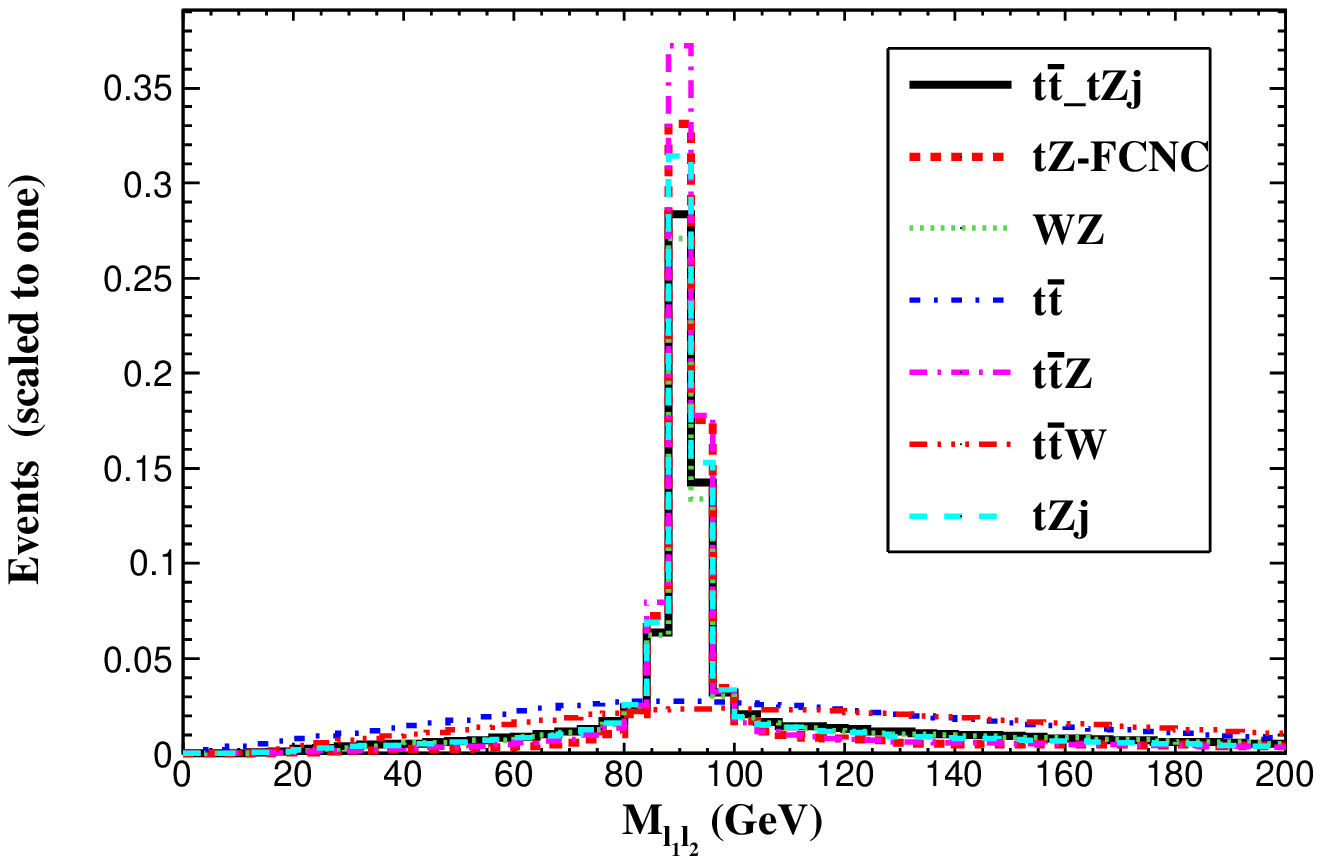}
\hspace{-2.0cm}\epsfxsize=9.5cm\epsffile{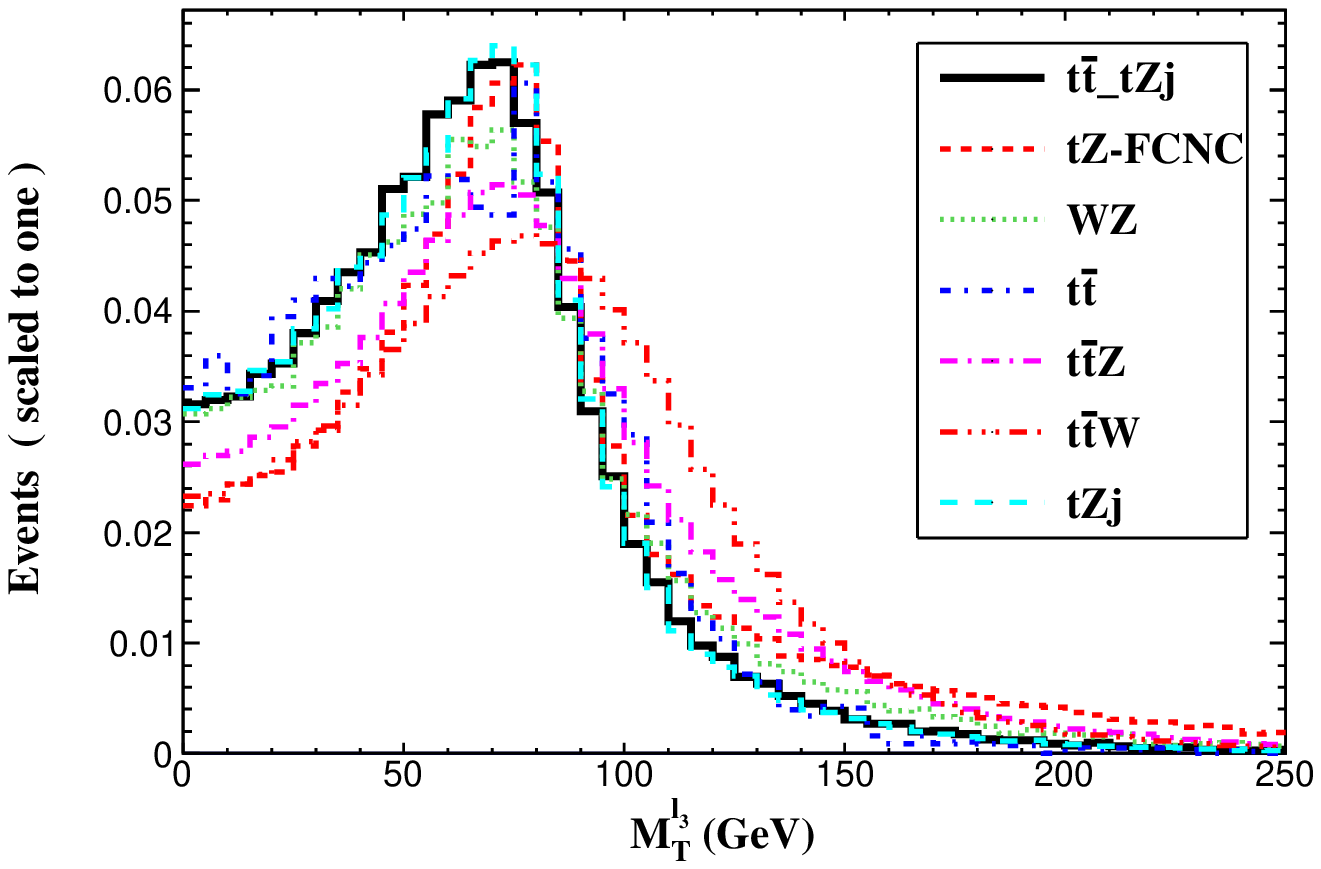}}
\centerline{\hspace{1.5cm}\epsfxsize=9.5cm\epsffile{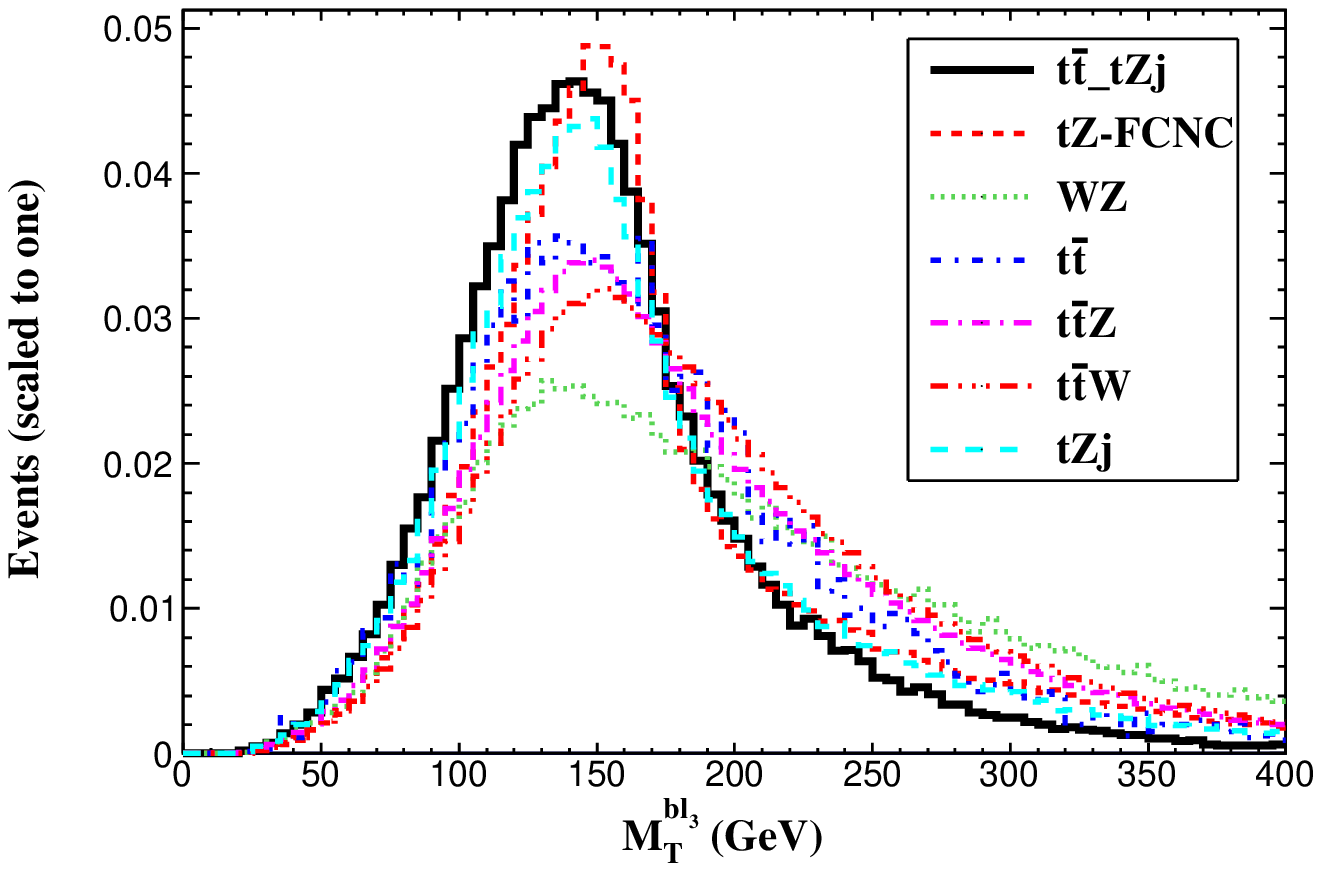}
\hspace{-2.0cm}\epsfxsize=9.5cm\epsffile{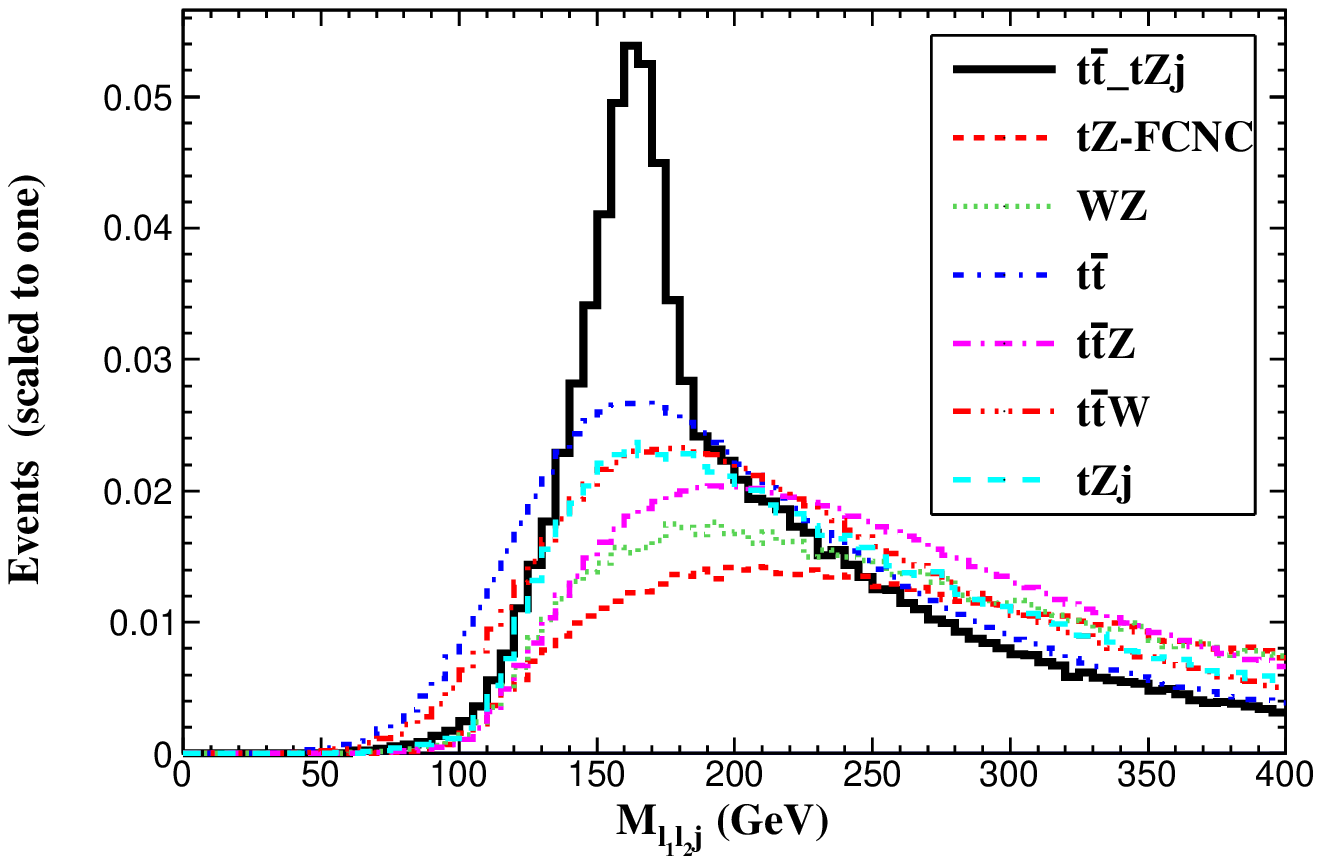}}
\caption{Normalised (to 1) distributions for the signals and SM backgrounds at the HL-LHC for Case A.}
\label{distribution1}
\end{center}
\end{figure}

For case A, the trilepton analysis aims to select $t\bar{t}$ events where one of the top quarks decays via the FCNC process ($t\to qZ\to q\ell_{1}\ell_{2}$) while the other top quark decays leptonically ($t\to Wb\to \ell_{3}\nu b$). Here, the leptons $\ell_{1}$ and $\ell_{2}$ are the two Opposite-Sign and Same-Flavour~(OSSF) leptons that are assumed to be the product of the $Z$-boson decay, whereas the third lepton, $\ell_{3}$, is assumed to originate from the leptonically decaying top quark, with the $b$-tagged jet emerging from the $t\to b W^+$ decay and the light jet $j$ is the non-$b$-tagged one. Therefore, the
following preselection is used for Case A~(Cut 1):
 \begin{itemize}
\item
exactly three isolated leptons with $p_{T}>30 \rm ~GeV$, in which at least one OSSF lepton pair;
\item
at least two jets with $p_{T}>40 \rm ~GeV$, with exactly one of them $b$-tagged;
\item
the missing transverse energy $E_{T}^{\rm miss}>30 ~\rm GeV$.
\end{itemize}

In Fig.~\ref{distribution1}, we plot some differential distributions  for  signals and SM backgrounds at the HL-LHC, such as the invariant mass distributions of the two leptons, $M_{\ell_{1}\ell_{2}}$, the transverse mass distribution  for $M_{T}(\ell_{3})$ and $M_{T}( b\ell_{3})$ as well as the triple  invariant mass, $M_{\ell_{1}\ell_{2} j}$. Furthermore, the top quark transverse cluster mass can be defined as~\cite{Conte:2014zja}
\beq
M_T^{2}\equiv(\sqrt{(p_{\ell_{3}}+p_{b})^{2}+|\vec{p}_{T,\ell_{3}}+\vec{p}_{T,b}|^{2}}+|\vec{\slashed p}_T| )^{2}-|\vec{p}_{T,\ell_{3}}+\vec{p}_{T,b}+\vec{\slashed p}_T|^{2},
\eeq
where $\vec{p}_{T,\ell_{3}}$ and  $\vec{p}_{T,b}$ are the transverse momenta of the third charged lepton and $b$-quark, respectively, and $\vec{\slashed p}_T$ is the missing transverse momentum determined by the negative sum of the visible momenta in the transverse direction.

According to the above analysis, we can impose the following set of cuts.
 \begin{itemize}
\item
(Cut 2) Two of the same-flavour leptons in each event are required to have
opposite electric charge and have an invariant mass, $M_{\ell_{1}\ell_{2}}$, compatible with the $Z$ boson mass, i.e.,
$|M(\ell_{1}\ell_{2})-m_{Z}|< 15 \rm ~GeV$.
\item
(Cut 3)  The transverse mass of the $W^\pm$ candidate is required to be $50~{\rm GeV} < M_T^{\ell_{3}} < 100~{\rm GeV}$ whereas the transverse mass of the top quark is required to be $100 ~{\rm GeV} < M_T^{b\ell_{3}} < 200~{\rm GeV}$.
\item
(Cut 4) The triple invariant mass $M_{\ell_{1}\ell_{2} j}$ cut is such that $140 ~{\rm GeV} < M_{\ell_{1}\ell_{2} j} < 200 ~{\rm GeV}$.
\end{itemize}

\begin{table}[ht!]
\fontsize{12pt}{8pt}\selectfont \caption{The cut flow of the cross sections (in fb) for the signals and SM backgrounds at the HL-LHC with $\kappa_{tuZ}=\lambda_{tuZ}=0.1$ and $\kappa_{tcZ}=\lambda_{tcZ}=0.1$ (in the brackets) for Case A. \label{cutflow1A}}
\begin{center}
\newcolumntype{C}[1]{>{\centering\let\newline\\\arraybackslash\hspace{0pt}}m{#1}}
{\renewcommand{\arraystretch}{1.5}
\scalebox{0.9}{
\begin{tabular}{C{1.4cm}| C{2.1cm}| C{2.1cm}| C{2.3cm} |C{2.3cm} |C{1.2cm}C{1.3cm}C{1.3cm}C{1.3cm} C{1.3cm} }
\hline
 \multirow{3}{*}{Cuts}& \multicolumn{4}{c|}{Signals}&\multicolumn{5}{c}{Backgrounds} \\ \cline{2-10}
&\multicolumn{2}{c|}{$t\bar{t}\to tZj$} &\multicolumn{2}{c|}{ $pp\to tZ$ } & \multirow{2}{*}{$WZ$} &\multirow{2}{*}{$t\bar{t}$ }& \multirow{2}{*}{$t\bar{t}Z$ } & \multirow{2}{*}{$t\bar{t}W$}& \multirow{2}{*}{$tZj$}\\   \cline{2-5}
&$tZq~(\sigma^{\mu\nu})$ & $tZq~(\gamma^{\mu})$ &$tZq~(\sigma^{\mu\nu})$ & $tZq~(\gamma^{\mu})$ &  & & & & \\   \cline{1-10}\hline
Basic &31.8~(33.4)&23.1~(24.3)&44~(7.6)&10.1~(2.2) &5.22&24618&8.32&1.36&4.23\\
Cut 1 &5.9~(5.6)&4.3~(4.2)&7.18~(1.15)&1.34~(0.28)&0.86&1.36&0.49&0.097&0.55\\
Cut 2 &4.5~(4.3)&~3.41~(3.25)&5.94~(0.95)&1.09~(0.23)&0.64&0.25&0.37&0.012&0.43\\
Cut 3 &1.93~(1.8)&1.46~(1.36)&2.39~(0.41)&0.47~(0.1)&0.14&0.085&0.12&0.0034&0.18\\
Cut 4 &0.91~(0.81)&0.68~(0.61)&0.2~(0.046)&0.077~(0.018)&0.031&0.027&0.028&0.0015&0.048
\\
\hline
\end{tabular} }}
 \end{center}
 \end{table}

\begin{table}[ht!]
\fontsize{12pt}{8pt}\selectfont \caption{The cut flow of the cross sections (in fb) for the signals and SM backgrounds at the HE-LHC with $\kappa_{tuZ}=\lambda_{tuZ}=0.1$ and $\kappa_{tcZ}=\lambda_{tcZ}=0.1$ (in the brackets) for Case A. \label{cutflow2A}}
\begin{center}
\newcolumntype{C}[1]{>{\centering\let\newline\\\arraybackslash\hspace{0pt}}m{#1}}
{\renewcommand{\arraystretch}{1.5}
\scalebox{0.9}{
\begin{tabular}{C{1.4cm}| C{2.1cm}| C{2.1cm}| C{2.3cm} |C{2.3cm} |C{1.2cm}C{1.3cm}C{1.3cm}C{1.3cm} C{1.3cm} }
\hline
 \multirow{3}{*}{Cuts}& \multicolumn{4}{c|}{Signals}&\multicolumn{5}{c}{Backgrounds} \\ \cline{2-10}
&\multicolumn{2}{c|}{$t\bar{t}\to tZj$} &\multicolumn{2}{c|}{ $pp\to tZ$ } & \multirow{2}{*}{$WZ$} &\multirow{2}{*}{$t\bar{t}$ }& \multirow{2}{*}{$t\bar{t}Z$ } & \multirow{2}{*}{$t\bar{t}W$}& \multirow{2}{*}{$tZj$}\\   \cline{2-5}
&$tZq~(\sigma^{\mu\nu})$ & $tZq~(\gamma^{\mu})$ &$tZq~(\sigma^{\mu\nu})$ & $tZq~(\gamma^{\mu})$ &  & & & & \\   \cline{1-10}\hline
Basic&179~(188)&129~(135)&170~(39)&35~(10) &13.5&71187&42&4.8&15.4\\
Cut 1 &29~(28)&22~(21)&27~(5.8)&4.6~(1.26)&2.73&4.88&2.67&0.35&1.93\\
Cut 2 &22~(21)&17~(16)&22.5~(4.8)&3.7~(1.0)&2.04&0.92&1.97&0.038&1.51\\
Cut 3 &9.1~(8.64)&7.0~(6.5)&8.16~(1.86)&1.49~(0.42)&0.41&0.31&0.58&0.011&0.59\\
Cut 4 &4.1~(3.9)&3.11~(2.75)&0.63~(0.18)&0.21~(0.072)&0.087&0.043&0.12&0.0048&0.144
\\
\hline
\end{tabular} }}
 \end{center}
 \end{table}

\begin{table}[ht!]
\fontsize{12pt}{8pt}\selectfont \caption{The cut flow of the cross sections (in fb) for the signals and SM backgrounds at the FCC-hh with $\kappa_{tuZ}=\lambda_{tuZ}=0.1$ and $\kappa_{tcZ}=\lambda_{tcZ}=0.1$ (in the brackets) for Case A. \label{cutflow3A}}
\begin{center}
\newcolumntype{C}[1]{>{\centering\let\newline\\\arraybackslash\hspace{0pt}}m{#1}}
{\renewcommand{\arraystretch}{1.5}
\scalebox{0.9}{
\begin{tabular}{C{1.4cm}| C{2.1cm}| C{2.1cm}| C{2.3cm} |C{2.3cm} |C{1.2cm}C{1.3cm}C{1.2cm}C{1.2cm} C{1.2cm} }
\hline
 \multirow{3}{*}{Cuts}& \multicolumn{4}{c|}{Signals}&\multicolumn{5}{c}{Backgrounds} \\ \cline{2-10}
&\multicolumn{2}{c|}{$t\bar{t}\to tZj$} &\multicolumn{2}{c|}{ $pp\to tZ$ } & \multirow{2}{*}{$WZ$} &\multirow{2}{*}{$t\bar{t}$ }& \multirow{2}{*}{$t\bar{t}Z$ } & \multirow{2}{*}{$t\bar{t}W$}& \multirow{2}{*}{$tZj$}\\   \cline{2-5}
&$tZq~(\sigma^{\mu\nu})$ & $tZq~(\gamma^{\mu})$ &$tZq~(\sigma^{\mu\nu})$ & $tZq~(\gamma^{\mu})$ &  & & & & \\   \cline{1-10}\hline
Basic&2135~(2315)&1532~(1662)&1122~(455)&290~(127) &267&764935&351&46&155\\
Cut 1 &440~(377)&335~(279)&276~(98)&56~(21)&61&60&22&5.6&31\\
Cut 2 &330~(280)&102~(86)&224~(80)&44~(17)&45&9.5&17&0.53&24\\
Cut 3 &134~(109)&102~(86)&85.2~(31.2)&17.6~(7.3)&8.7&3.7&4.9&0.14&8.7\\
Cut 4 &70~(57)&54.4~(43.5)&9.03~(4.01)&3.32~(1.53)&2.01&1.07&0.82&0.07&1.83
\\
\hline
\end{tabular} }}
 \end{center}
 \end{table}

We use the same selection cuts for the HE-LHC and FCC-hh
analysis because the distributions are very similar to the case of the  HL-LHC.
The effects of the described  cuts  on the signal and SM  background processes are illustrated in Tabs.~\ref{cutflow1A}--\ref{cutflow3A}.
Due to the different $b$-tagging  rates for $u$- and $c$-quarks, we give the events separately for $q = u, c$ for the signals. One can see that, at the end of the cut flow, the largest SM  background is the $pp\to tZj$ process, which is about 0.048 fb, 0.144 fb and 1.45 fb at the HL-LHC, HE-LHC and FCC-hh, respectively. Besides, the $W^\pm Z$ + jets and  $t\bar{t}Z$ processes can also generate significant contributions to the SM background. Obviously, the dominant signal contribution comes from the $t\bar{t}$-FCNC  process, but the contribution from the $tZ$-FCNC production process cannot be ignored, especially for the $tuZ$ couplings.

\subsection{The selection cuts for Case B}
For this case, we will mainly concentrate on the signal from the $ug\to tZ$ process due to the relative large cross section and we will veto extra jets in the following analysis. However, the final signals for Case A could also be considered as a source for Case B if the light quark is missed by the detector. Hence, we combine these processes into the complete signal events.

\begin{figure}[htb]
\begin{center}
\centerline{\hspace{1.5cm}\epsfxsize=9.5cm\epsffile{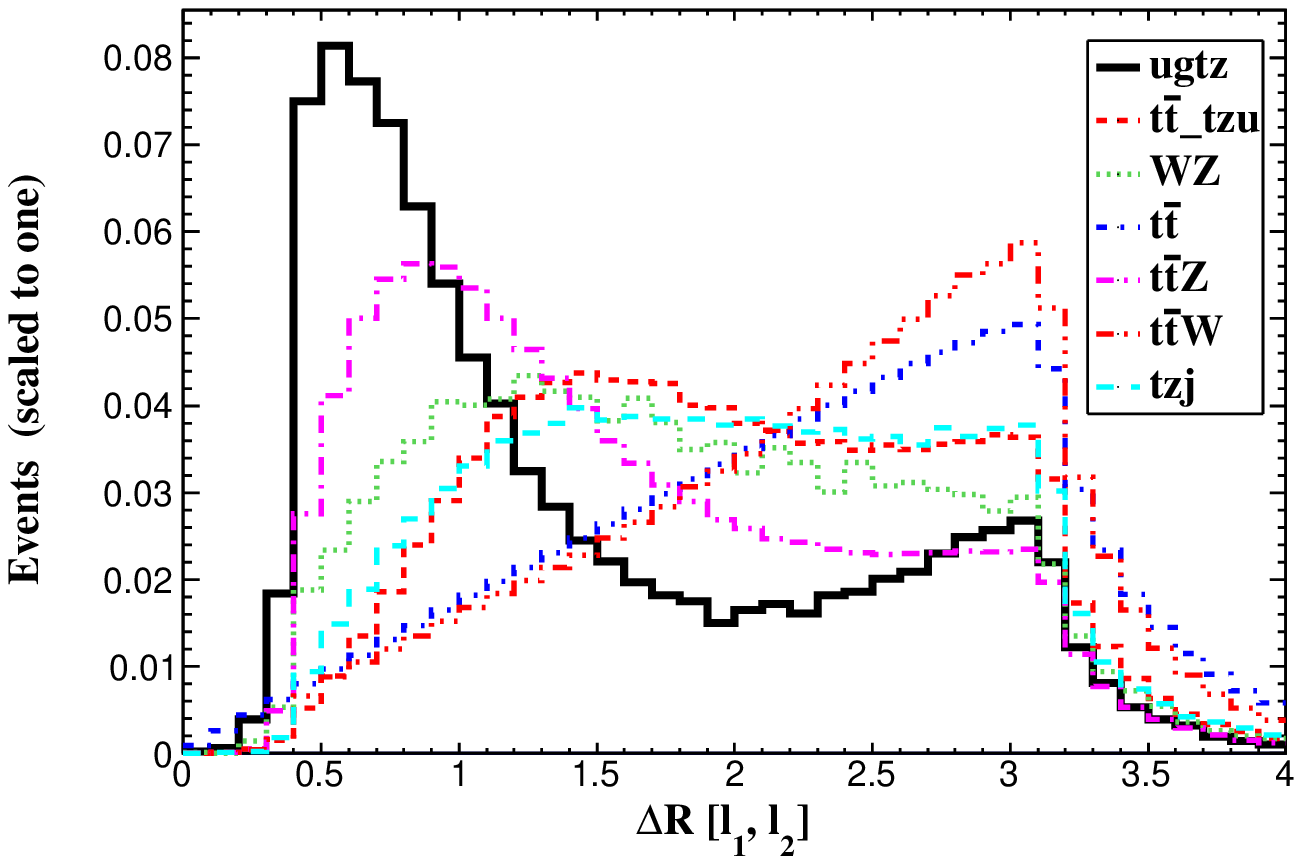}
\hspace{-2.0cm}\epsfxsize=9.5cm\epsffile{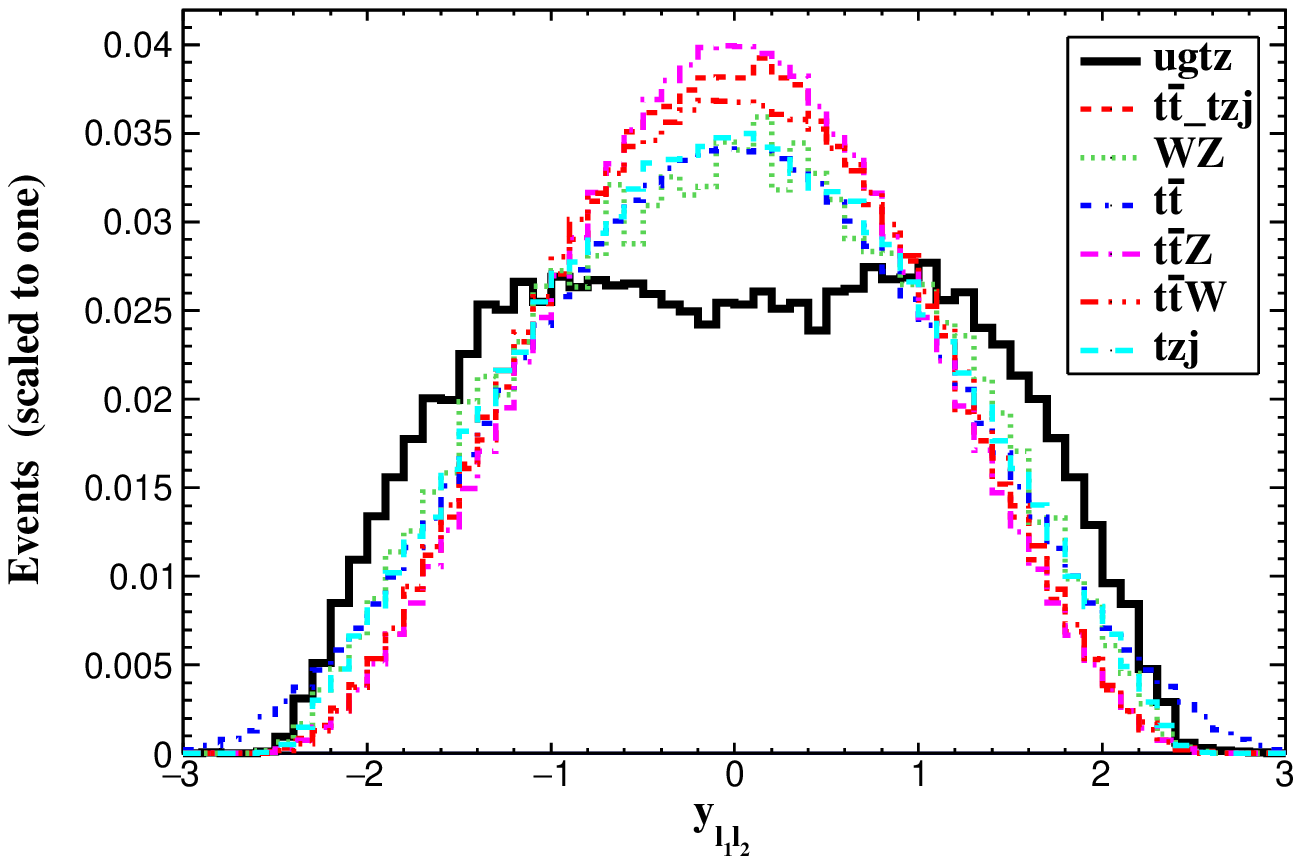}}
\caption{Normalised (to 1) distributions for the signals and SM backgrounds at the HL-LHC for Case B.}
\label{distribution2}
\end{center}
\end{figure}

The process  $ug\to tZ$ should include two leptons with positive charge, one coming from the decay $Z\to \ell^{+}\ell^{-}$ and the other from the top quark decay $t\to W^{+}b\to \ell^{+}\nu b$.
Because the distributions for the signal and backgrounds  are similar for the invariant mass $M_{\ell_1\ell_2}$ as well as the transverse masses $M_{T}(\ell_{3})$ and $M_{T}( b\ell_{3})$, we only plot the distributions for the distance of the OSSF lepton pair, $\Delta R(\ell_{1},\ell_{2})$, and the rapidity of the OSSF lepton pair, $y_{\ell_{1}\ell_{2}}$,  in Fig.~\ref{distribution2}. (Here, the distributions are obtained at the HL-LHC, but the pattern is very similar at the HE-LHC and FCC-hh.)  One can see that, for Case B, the $Z$ boson from the $ug\to tZ$ process concentrates in the forward and backward regions since the partonic c.m. frame is highly boosted along the direction of the $u$-quark.

Thus, we can impose the following set of cuts for Case B.
 \begin{itemize}
\item
(Cut 1) There are three leptons in which at least two  with positive charge and $p_{T}>30 \rm ~GeV$, exactly one $b$-tagged jet with $p_{T}>40 \rm ~GeV$ and the event is rejected if the $p_{T}$ of the subleading jet is greater than 25~GeV.
\item
(Cut 2) The distance between the OSSF lepton pair should lie within $\Delta R(\ell_{1},\ell_{2})\in [0.4, 1.2]$ while the corresponding invariant mass  is  required to be $|M(\ell_{1}\ell_{2})-m_{Z}|< 15 \rm ~GeV$.
\item
(Cut 3) The transverse masses of the reconstructed $W^\pm$ boson and top quark masses are required to satisfy $50 ~{\rm GeV} < M_T^{\ell_{3}} < 100 ~{\rm GeV}$ and $100 ~{\rm GeV} < M_T^{b\ell_{3}} < 200 ~{\rm GeV}$, respectively.
\item
(Cut 4) The rapidity of the OSSF lepton pair is required to be $|y_{\ell_{1}\ell_{2}}| > 1.0$.
\end{itemize}

The effects of these cuts on the signal and
background processes for Case B are illustrated in Tabs. \ref{cutflow1B}--\ref{cutflow3B}. One can see that all  backgrounds can be suppressed efficiently after imposing such a selection. At the end of the cut flow, the $W^\pm Z$ + jets and $t\bar{t}$ production processes are the dominant SM backgrounds mainly due to the initially large cross sections.

\begin{table}[ht!]
\fontsize{12pt}{8pt}\selectfont \caption{The cut flow of the cross sections (in $\times 10^{-2}$~fb) for the signals and SM backgrounds at the HL-LHC with $\kappa_{tuZ}=0.1$ and $\lambda_{tuZ}=0.1$ (in the brackets)  for Case B. \label{cutflow1B}}
\begin{center}
\newcolumntype{C}[1]{>{\centering\let\newline\\\arraybackslash\hspace{0pt}}m{#1}}
{\renewcommand{\arraystretch}{1.5}
\begin{tabular}{C{2.0cm}| C{2.2cm} |C{2.2cm} |C{1.4cm}C{1.7cm}C{1.4cm}C{1.4cm} C{1.4cm} }
\hline
 \multirow{2}{*}{Cuts}& \multicolumn{2}{c|}{Signals}&\multicolumn{5}{c}{Backgrounds} \\ \cline{2-8}
&$ug\to tZ$ & $t\bar{t}\to tZj$& $WZ$ &$t\bar{t}$ & $t\bar{t}Z$  &$t\bar{t}W$&$tZj$\\   \cline{2-8}\hline
Basic&3365~(856)&2664~(1926)&474&$2.2\times 10^{6}$&602&233&367\\
Cut 1 &319~(61)&23~(18)&14&38&1.2&4.5&1.3\\
Cut 2 &184~(23)&5.6~(4.3)&3.5&1.0&0.29&0.005&0.26\\
Cut 3 &108~(13.2)&3~(2.66)&0.9&0.43&0.07&0.01&0.14\\
Cut 4 &57~(7.2)&1.2~(1.1)&0.39&0.19&0.02&0.005&0.073
\\
\hline
\end{tabular} }
 \end{center}
 \end{table}

\begin{table}[ht!]
\fontsize{12pt}{8pt}\selectfont \caption{The cut flow of the cross sections (in fb) for the signals and SM backgrounds at the HE-LHC with $\kappa_{tuZ}=0.1$ and $\lambda_{tuZ}=0.1$ (in the brackets) for Case B. \label{cutflow2B}}
\begin{center}
\newcolumntype{C}[1]{>{\centering\let\newline\\\arraybackslash\hspace{0pt}}m{#1}}
{\renewcommand{\arraystretch}{1.5}
\begin{tabular}{C{2.0cm}| C{2.4cm} |C{2.4cm} |C{1.4cm}C{1.4cm}C{1.4cm}C{1.4cm} C{1.4cm} }
\hline
 \multirow{2}{*}{Cuts}& \multicolumn{2}{c|}{Signals}&\multicolumn{5}{c}{Backgrounds} \\ \cline{2-8}
&$ug\to tZ$ & $t\bar{t}\to tZj$   & $WZ$ &$t\bar{t}$ & $t\bar{t}Z$  & $t\bar{t}W$& $tZj$\\   \cline{1-8} \hline
Basic&123~(30)&153~(11)&14.2&64628&31.6&7.7&13.5\\
Cut 1 &7.9~(1.38)&1.0~(0.075)&0.31&1.05&0.04&0.12&0.043\\
Cut 2 &4.63~(0.54)&0.27~(0.018)&0.075&0.043&0.009&0.0014&0.0087\\
Cut 3 &2.68~(0.32)&0.18~(0.01)&0.016&0.037&0.0021&0.0004&0.0046\\
Cut 4 &1.68~(0.203)&0.07~(0.003)&0.0064&0.018&0.0007&0.0002&0.0024
\\
\hline
\end{tabular} }
 \end{center}
 \end{table}

\begin{table}[ht!]
\fontsize{12pt}{8pt}\selectfont \caption{The cut flow of the cross sections (in fb) for the signals and SM backgrounds at the FCC-hh with $\kappa_{tuZ}=0.1$ and $\lambda_{tuZ}=0.1$ (in the brackets) for Case B. \label{cutflow3B}}
\begin{center}
\newcolumntype{C}[1]{>{\centering\let\newline\\\arraybackslash\hspace{0pt}}m{#1}}
{\renewcommand{\arraystretch}{1.5}
\begin{tabular}{C{2.0cm}| C{2.3cm} |C{2.3cm} |C{1.4cm}C{1.4cm}C{1.4cm}C{1.4cm} C{1.4cm} }
\hline
 \multirow{2}{*}{Cuts}& \multicolumn{2}{c|}{Signals}&\multicolumn{5}{c}{Backgrounds} \\ \cline{2-8}
&$ug\to tZ$ & $t\bar{t}\to tZj$   & $WZ$ &$t\bar{t}$ & $t\bar{t}Z$  & $t\bar{t}W$& $tZj$\\   \cline{1-8} \hline
Basic&727~(224)&1518~(1219)&313&697297&242&43&132\\
Cut 1 &24~(5.1)&2.4~(2.0)&4.1&4.3&0.035&0.33&0.144\\
Cut 2 &13.5~(1.67)&0.66~(0.39)&0.85&0.098&0.007&0.003&0.025\\
Cut 3 &8.12~(1.0)&0.35~(0.27)&0.12&0.049&0.0006&0.0007&0.011\\
Cut 4 &5.94~(0.73)&0.23~(0.13)&0.071&0.025&0.0003&0.0004&0.0077
\\
\hline
\end{tabular} }
 \end{center}
 \end{table}

\subsection{The $95\%$ CL exclusion limits}
To estimate the exclusion significance, we use the following expression~\cite{Cowan:2010js}:
\be
    Z_\text{excl} =\sqrt{2\left[S-B\ln\left(\frac{B+S+x}{2B}\right)
  - \frac{1}{\delta^2 }\ln\left(\frac{B-S+x}{2B}\right)\right] -
  \left(B+S-x\right)\left(1+\frac{1}{\delta^2 B}\right)},
 \ee
with
 \beq
 x=\sqrt{(S+B)^2- 4 \delta^2 S B^2/(1+\delta^2 B)}.
  \eeq
 Here, $S$ and $B$ represent the total signal and SM background events, respectively. Furthermore, $\delta$ is the percentage systematic error on the SM
background estimate. Following Refs.~\cite{Cowan:2010js,Kling:2018xud}, we define the regions with $Z_\text{excl} \leq 1.645$ as those that can be excluded at 95\% CL.
In the case of $\delta \to 0$,  the above expressions  are simplified as
\be
 Z_\text{excl} = \sqrt{2[S-B\ln(1+S/B)]}.
\ee
Using the results from Case A and Case B, we combine the significance for $Br(t\to uZ)$ with two kinds of couplings,
\beq
Z_\text{comb}=\sqrt{Z_\text{A}+Z_\text{B} }
\eeq
while, for $Br(t\to cZ)$, we only use the results from Case A.
\begin{figure}[htbp]
\begin{center}
\vspace{-0.5cm}
\centerline{\epsfxsize=6.5cm \epsffile{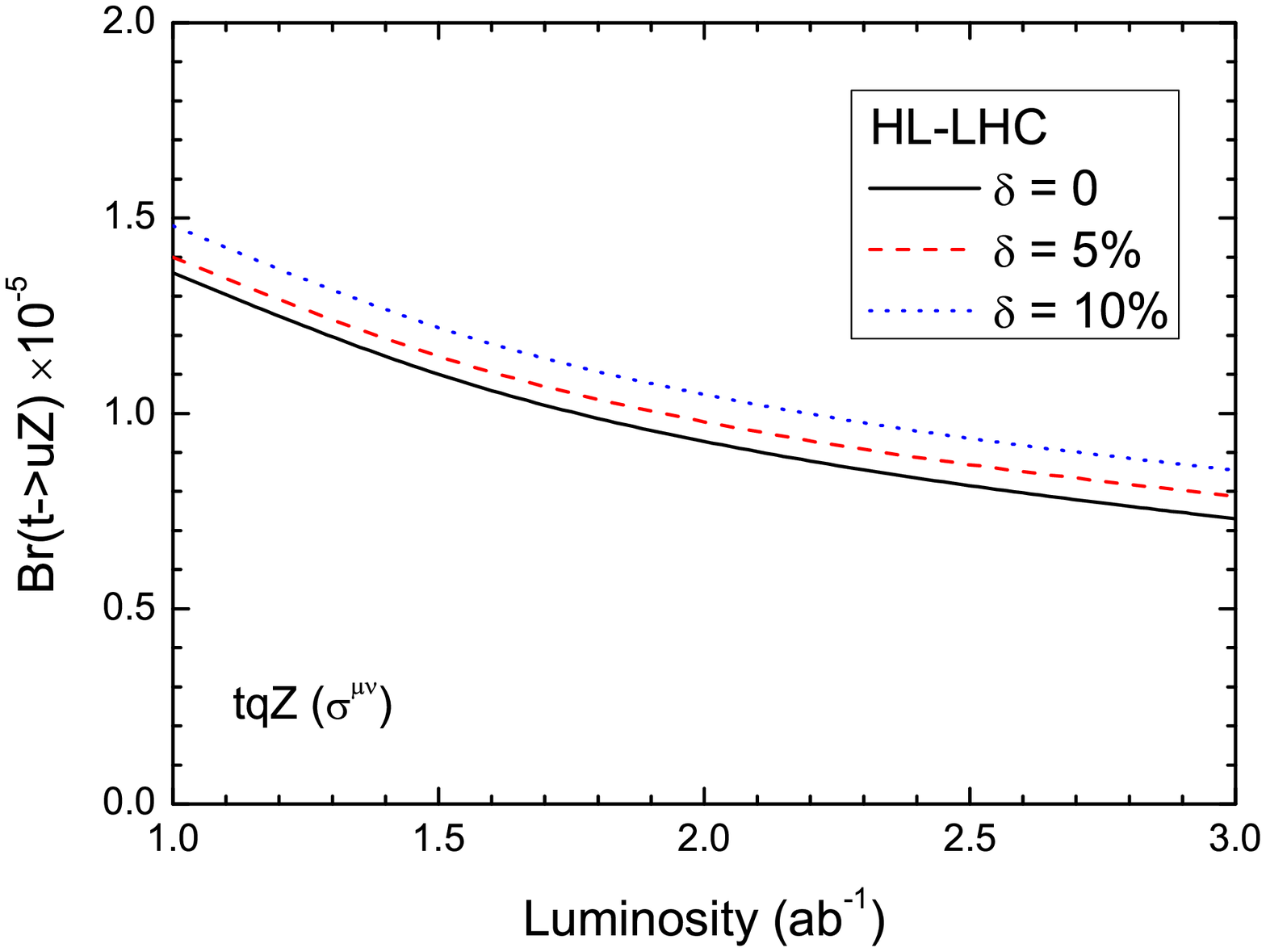}\hspace{-1.0cm}\epsfxsize=6.5cm \epsffile{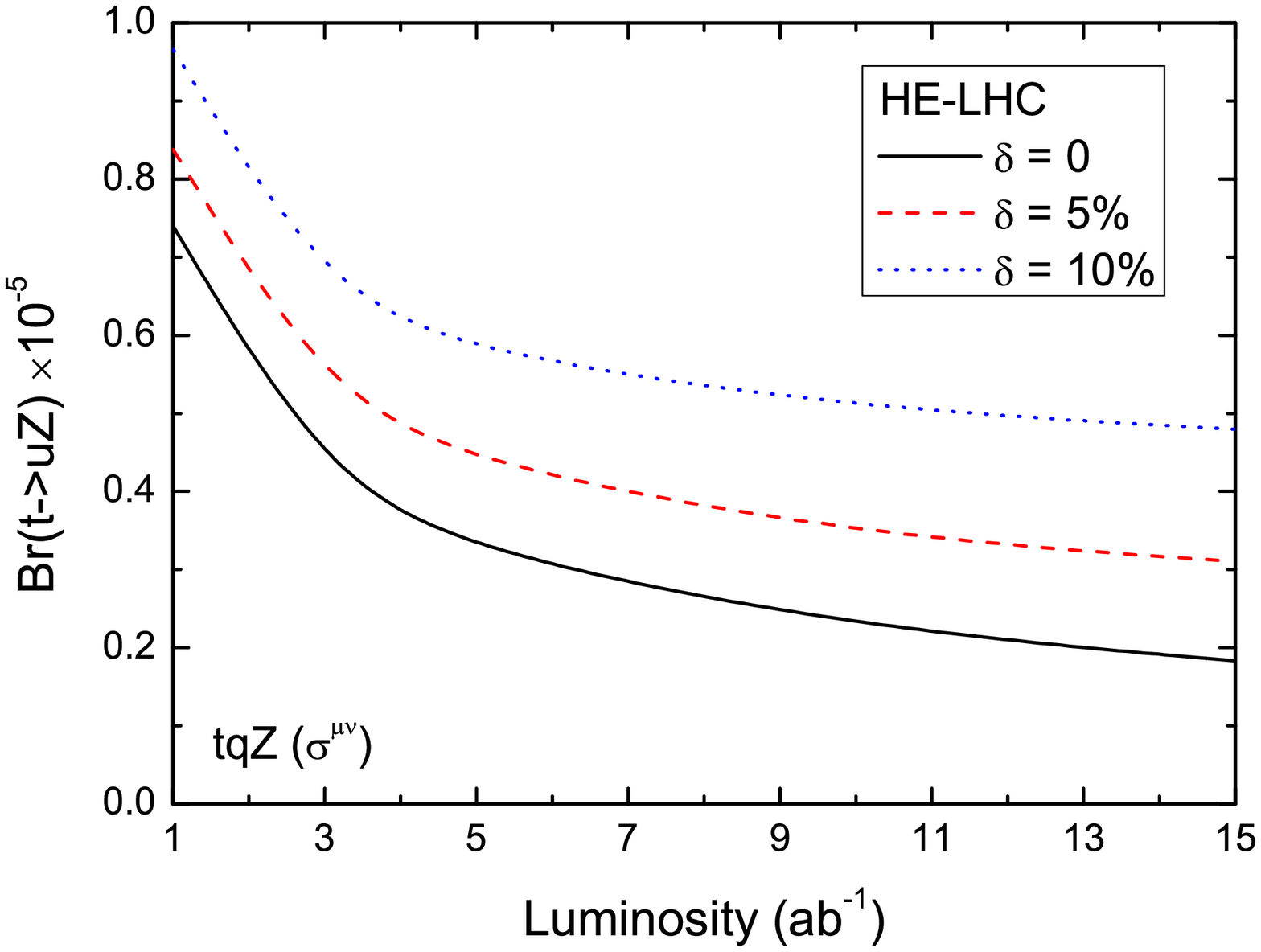}\hspace{-1.0cm}\epsfxsize=6.5cm \epsffile{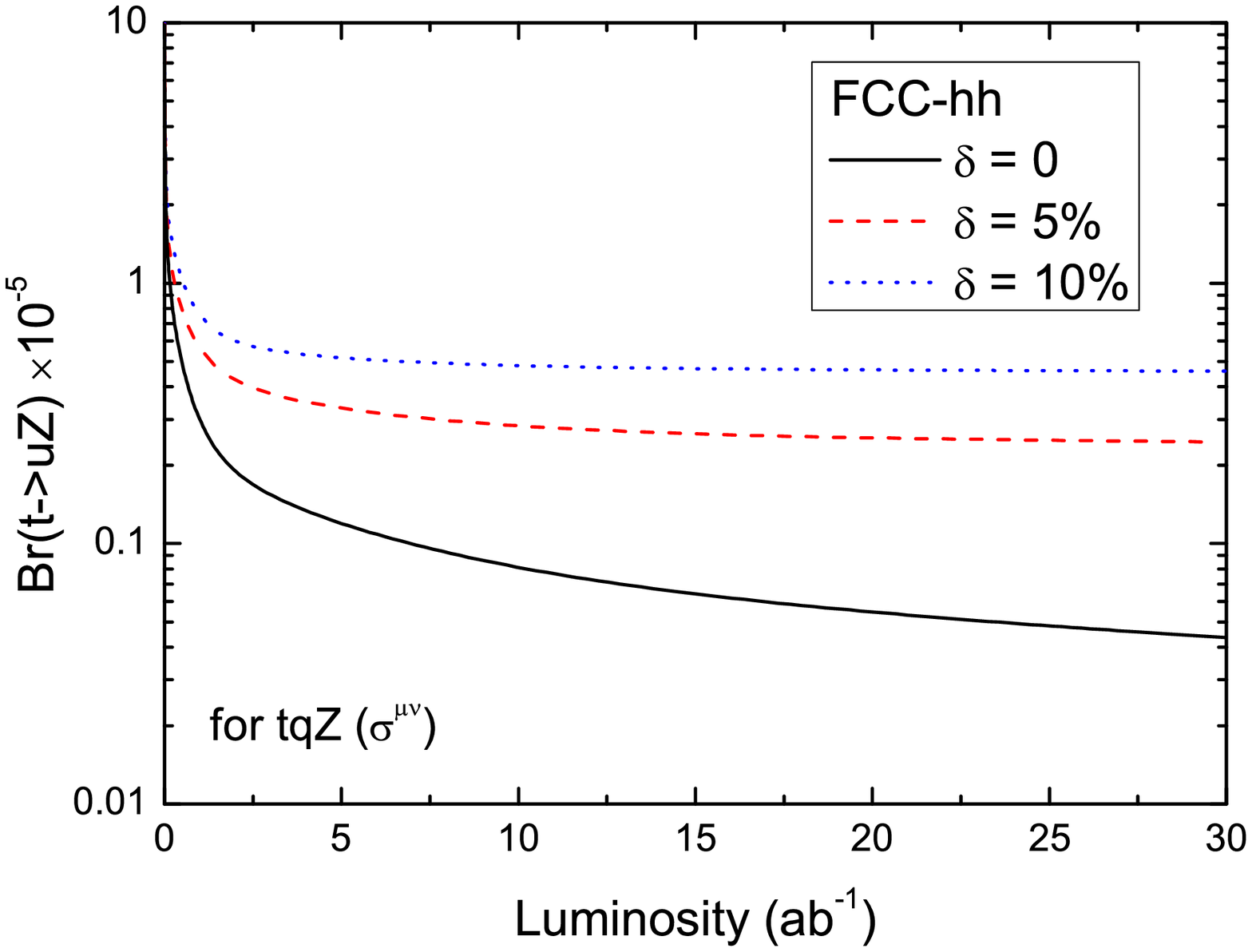}}
\centerline{\epsfxsize=6.5cm \epsffile{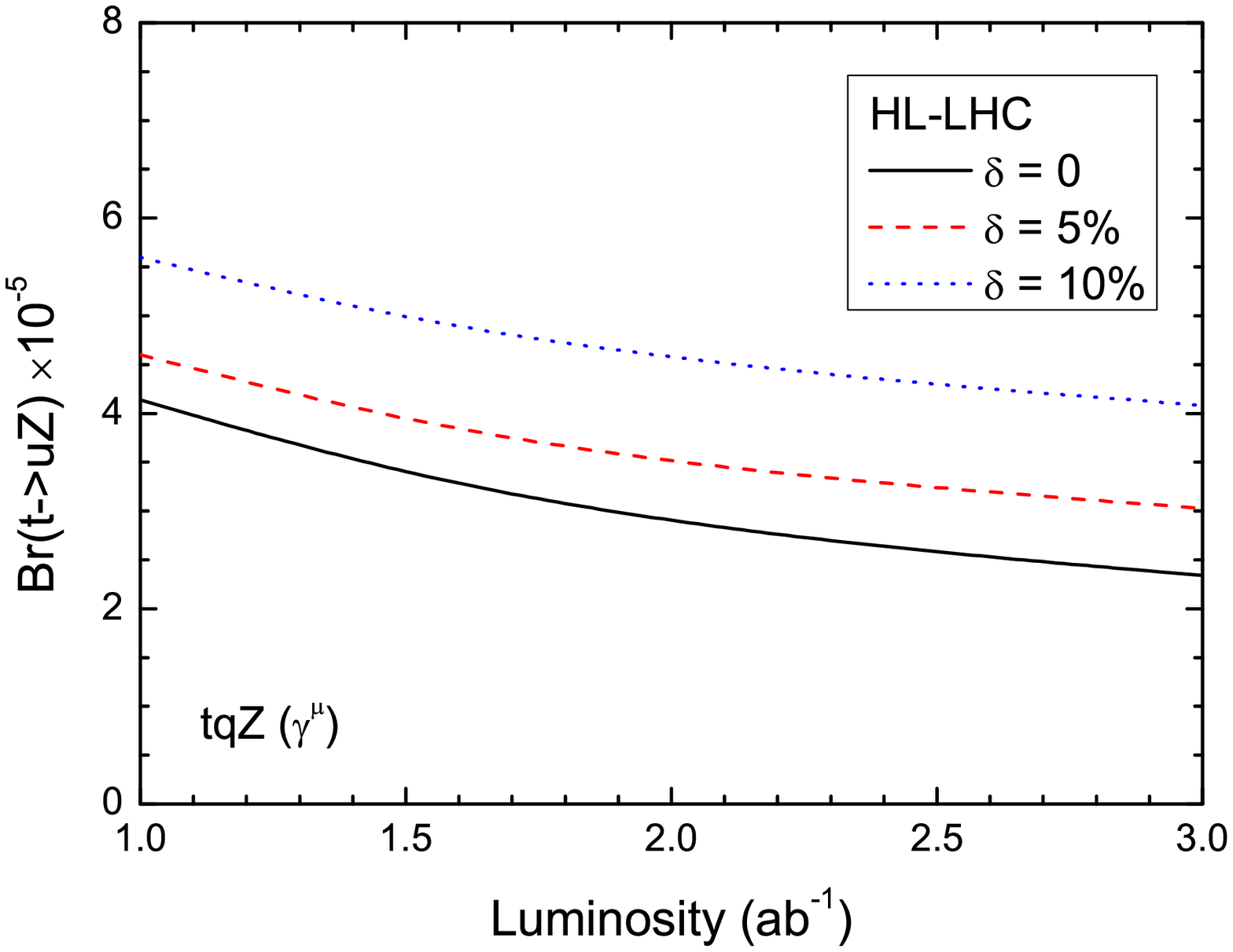}\hspace{-1.0cm}\epsfxsize=6.5cm \epsffile{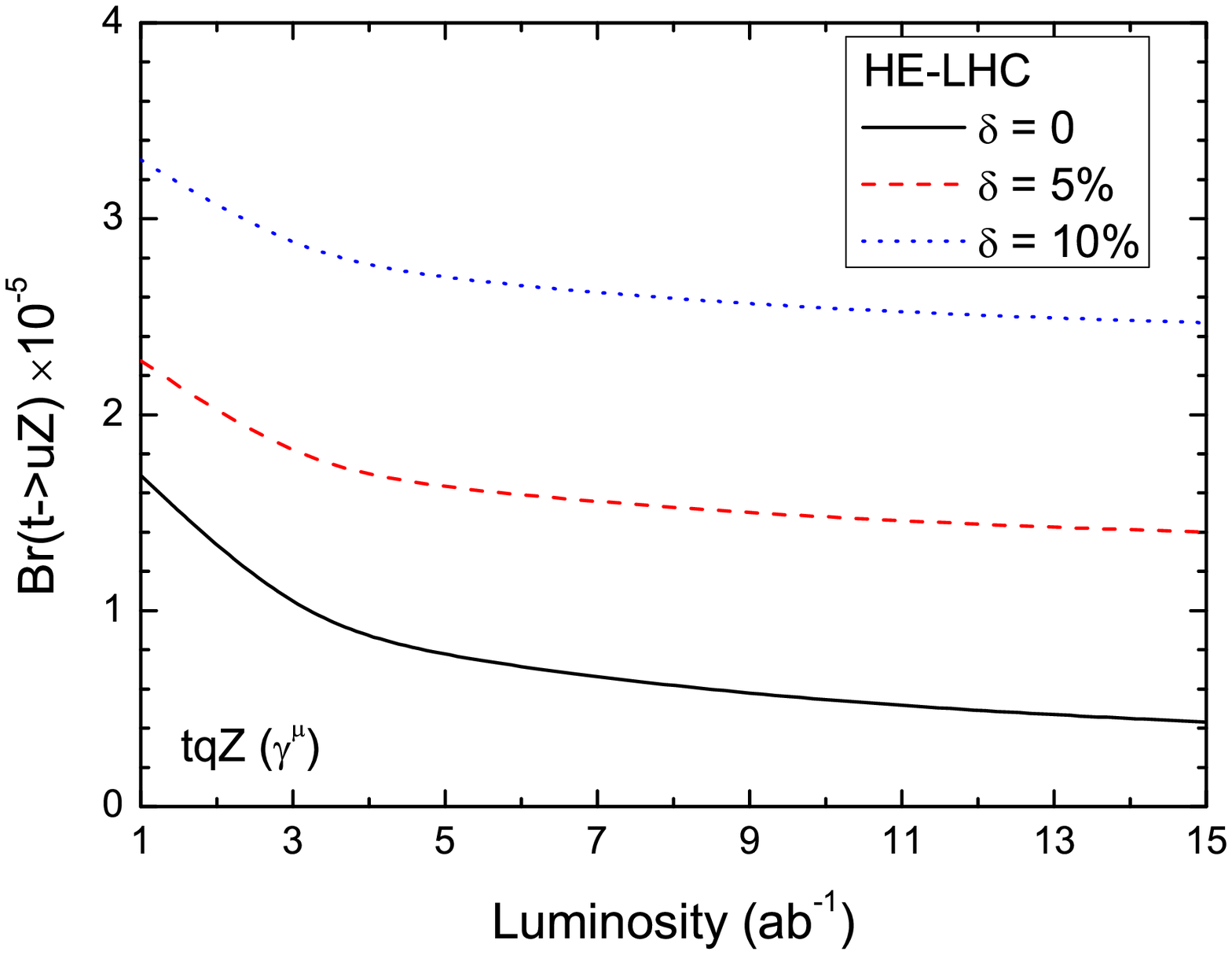}\hspace{-1.0cm}\epsfxsize=6.5cm \epsffile{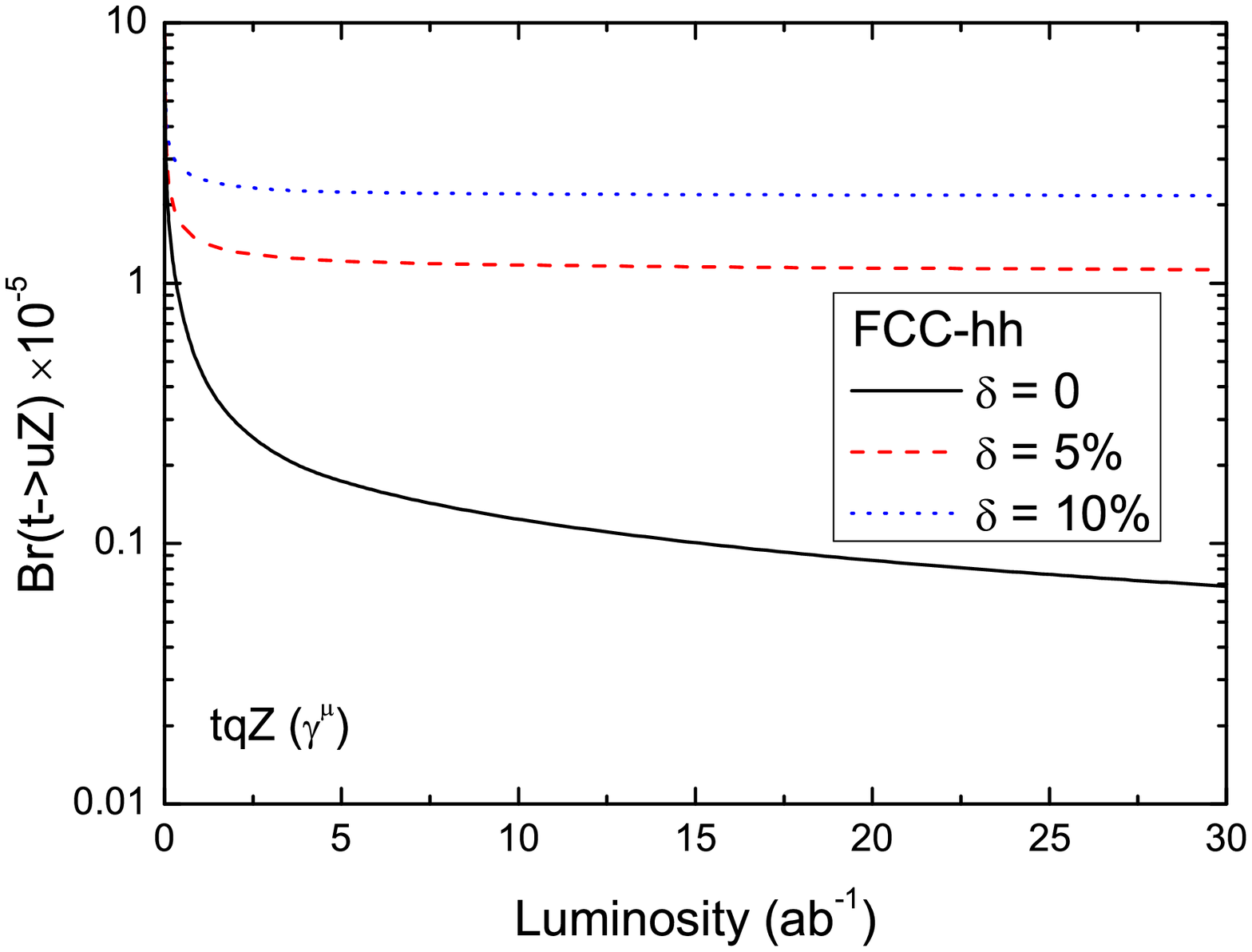}}
\caption{The combined $95\%$ CL contour plots in $L_{\rm int}-{\rm BR}(t\to uZ)$ planes for the tensor terms (upper) and the vector terms (below) at the HL-LHC (left), HE-LHC (middle) and FCC-hh (right). Three typical values for the systematic uncertainties, $\delta=0, 5\%, 10\%$, are taken.}
\label{tuz-A}
\end{center}
\end{figure}
\begin{figure}[htbp]
\begin{center}
\vspace{-0.5cm}
\centerline{\epsfxsize=6.5cm \epsffile{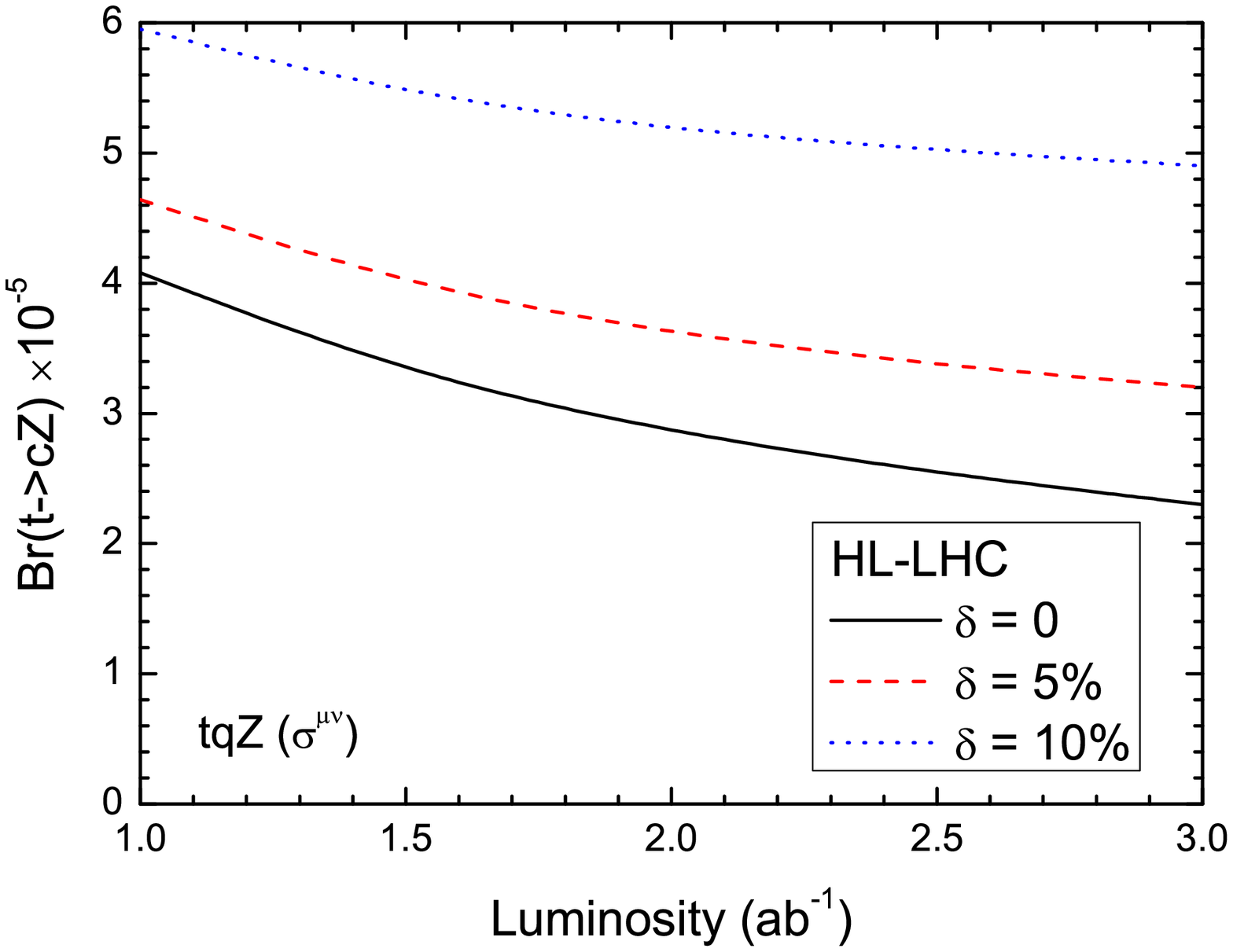}\hspace{-1.0cm}\epsfxsize=6.5cm \epsffile{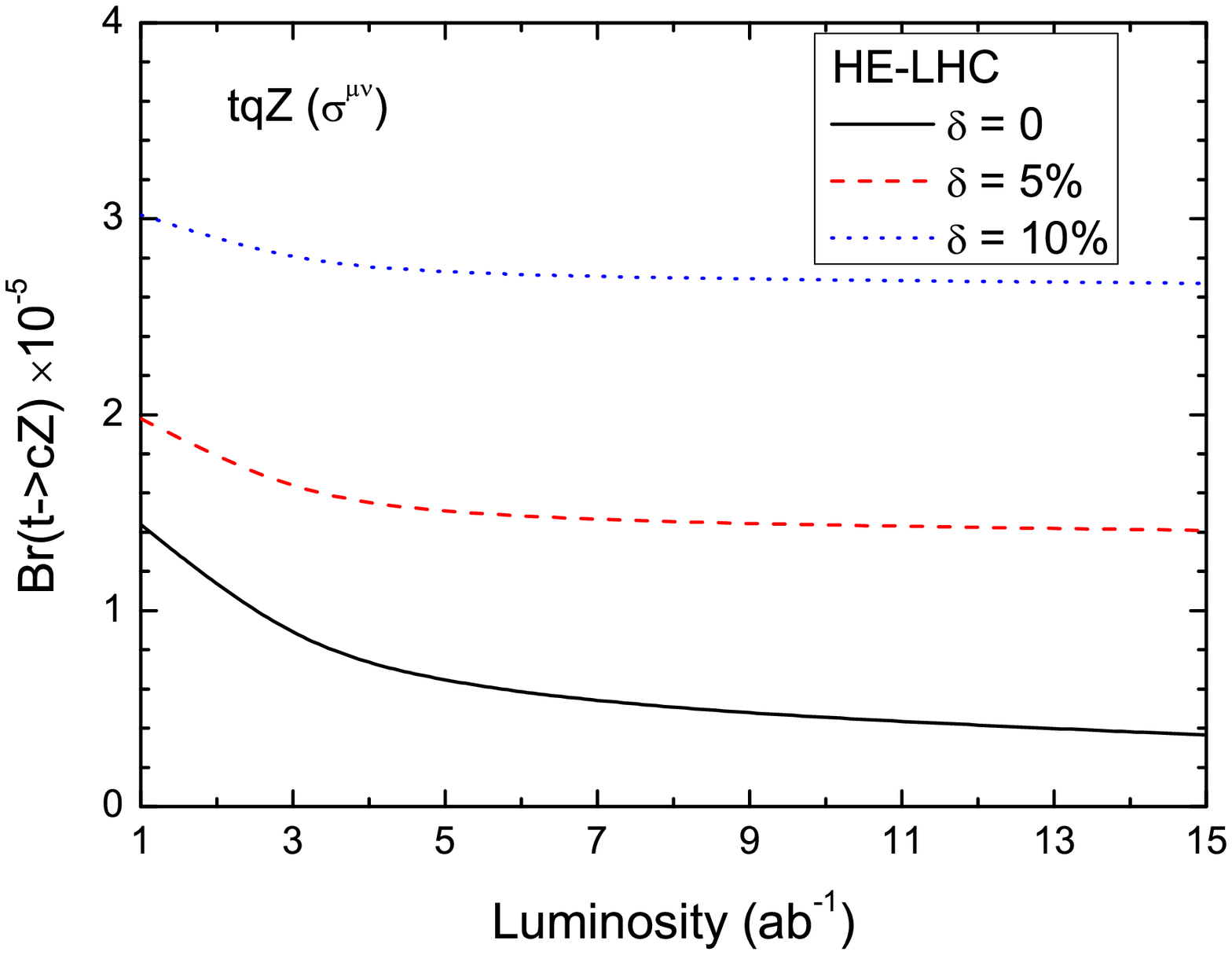}\hspace{-1.0cm}\epsfxsize=6.5cm \epsffile{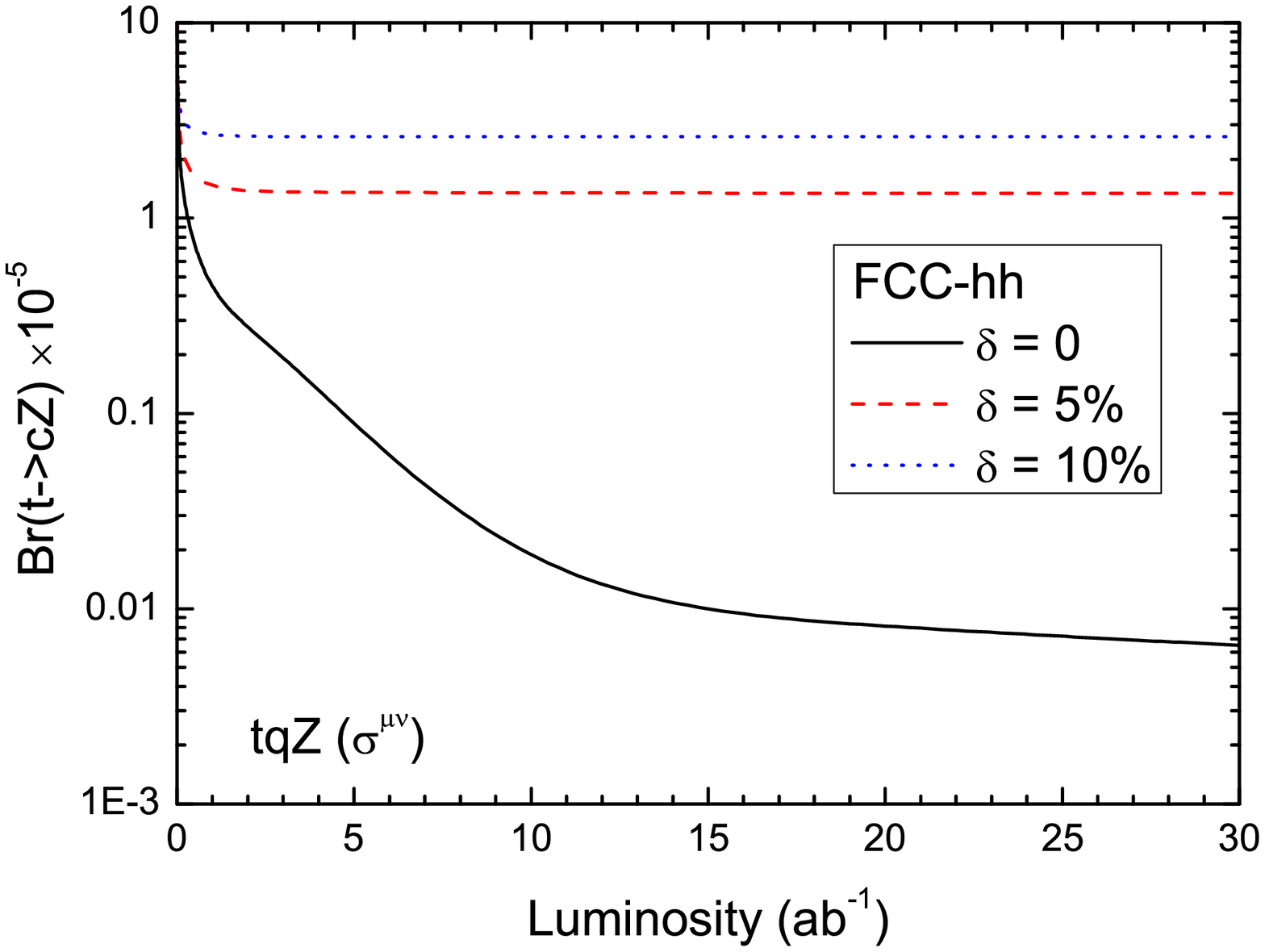}}
\centerline{\epsfxsize=6.5cm \epsffile{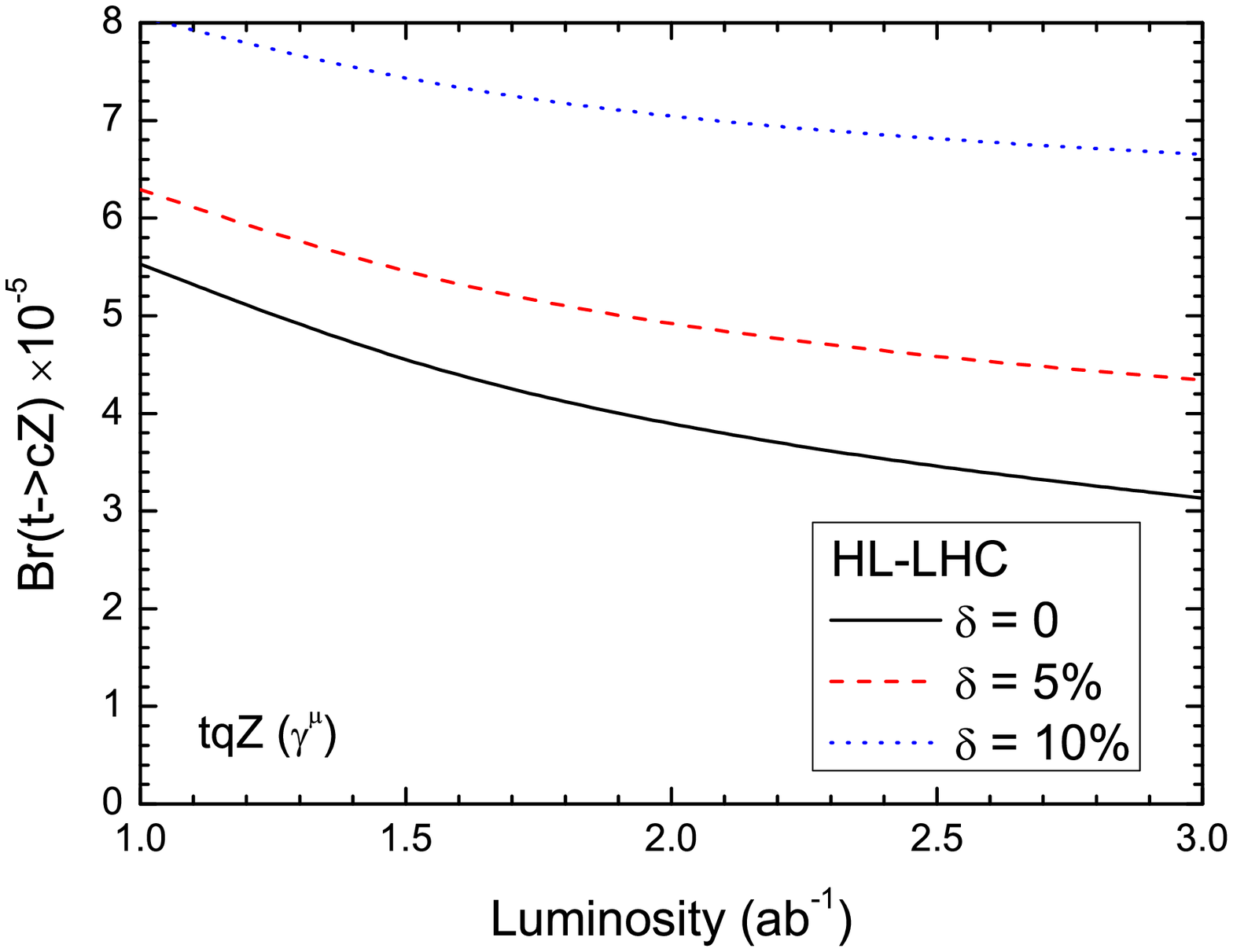}\hspace{-1.0cm}\epsfxsize=6.5cm \epsffile{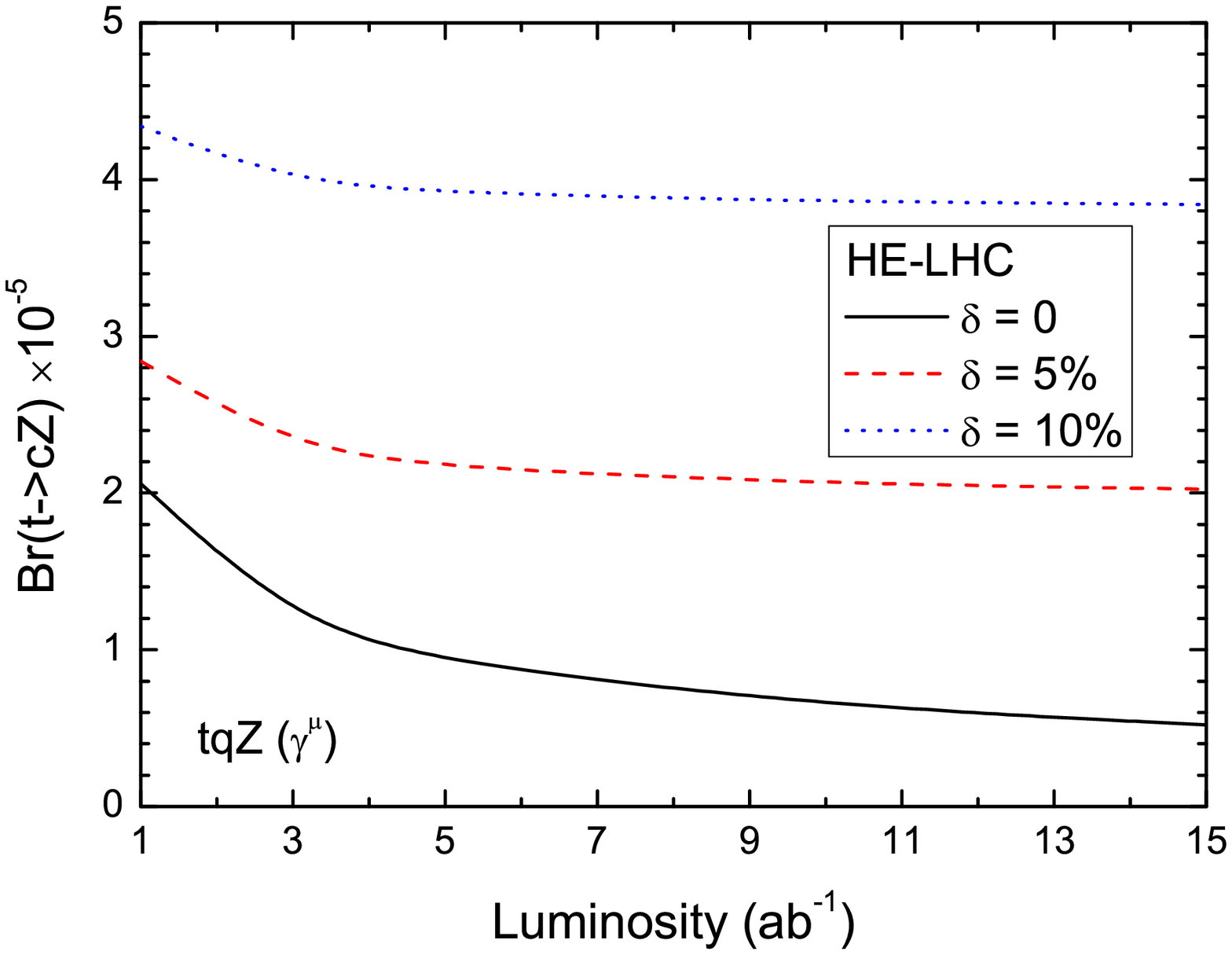}\hspace{-1.0cm}\epsfxsize=6.5cm \epsffile{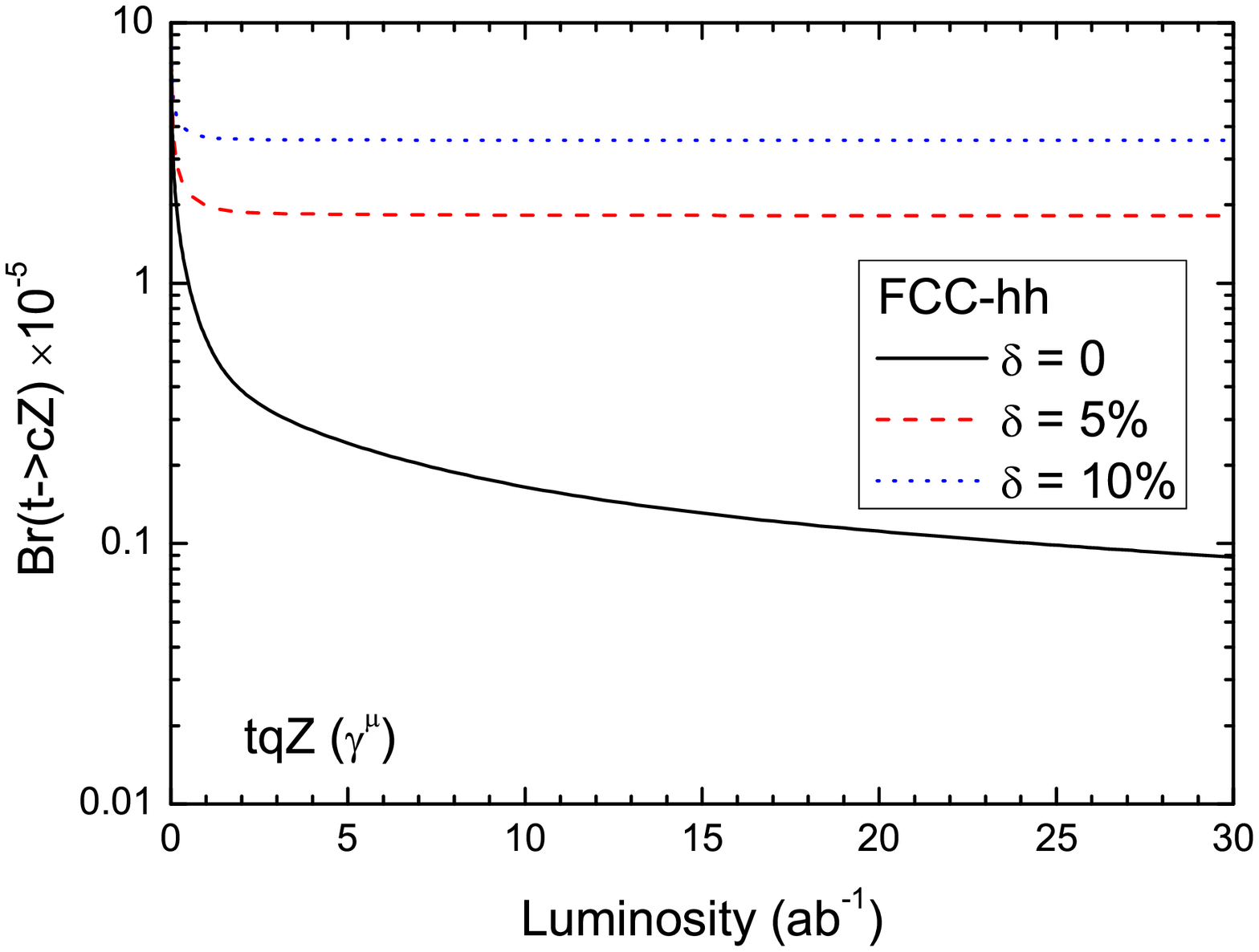}}
\caption{For Case A, the $95\%$ CL contour plots in $L_{\rm int}-{\rm BR}(t\to cZ)$ planes for the tensor terms (upper) and the vector terms (below) at the HL-LHC (left), HE-LHC (middle) and FCC-hh (right).  Three typical values for the systematic uncertainties, $\delta=0, 5\%, 10\%$, are taken.}
\label{tcz-A}
\end{center}
\end{figure}

In Figs.~\ref{tuz-A}--\ref{tcz-A},
 we plot the $95\%$ CL lines as a function of the integrated luminosity and BR$(t\to qZ)$ for the  two kinds of couplings with three typical values of systematic uncertainties: $\delta=0, 5\%$ and $10\%$.
 One can see from Fig.~\ref{tuz-A} that, for the tensor~(vector) terms, the combined $95\%$ CL  limits without systematic error on ${\rm BR}(t\to uZ)$ are $2.3~(5.3)\times 10^{-6}$ and $0.76~(1.2)\times 10^{-6}$ at the HE-LHC and FCC-hh with an integrated luminosity of 10 ab$^{-1}$, respectively. For this value of integrated luminosity, by taking
into account a $5\%$ systematic error, the obtained limits are about  $0.34~(1.47)\times 10^{-5}$ and $0.27~(1.21)\times 10^{-5}$, respectively, while, for the case
$\delta=10\%$, the $95\%$ CL limits on ${\rm BR}(t\to uZ)$  change to $0.51~(2.53)\times 10^{-5}$ and $0.48~(2.2)\times 10^{-5}$, respectively. From Fig.~\ref{tcz-A}, one can see that for the Case A, the $95\%$ CL  limits without systematic error on ${\rm BR}(t\to cZ)$ are $0.45~(0.64)\times 10^{-5}$ and $1.13~(1.54)\times 10^{-6}$ at the HE-LHC and FCC-hh with an integrated luminosity of 10 ab$^{-1}$, respectively. By taking
into account a $5\%$ systematic error, the obtained limits are about  $1.43~(2.06)\times 10^{-5}$ and $1.35~(1.82)\times 10^{-5}$, respectively.

\begin{table*}[htbp]
\begin{center}
\caption{The upper limits on BR$(t\to u(c)Z)$ at 95\% CL obtained at the HL-LHC, HE-LHC and FCC-hh. We consider systematic errors of 0\% and 10\% on the SM background events. \label{limit}}
\vspace{0.2cm}
\begin{tabular}{c| c| c| c |c |c |c}  \hline
\multirow{2}{*}{Branching fraction}& \multicolumn{2}{c|}{HL-LHC, 3 ab$^{-1}$} & \multicolumn{2}{c|}{HE-LHC, 15 ab$^{-1}$}&\multicolumn{2}{c}{FCC-hh, 30 ab$^{-1}$} \\ \cline{2-7}
 & $\delta=0$  & $\delta=10\%$& $\delta=0$  & $\delta=10\%$& $\delta=0$  & $\delta=10\%$ \\   \cline{2-7} \hline
BR$(t \to u Z)~(\sigma^{\mu\nu})$  &   $0.73 \times 10^{-5}$  &   $0.85\times 10^{-5}$ &   $1.83 \times 10^{-6}$ &   $4.8 \times 10^{-6}$&     $4.35\times 10^{-7}$&     $4.6 \times 10^{-6}$	\\
BR$(t \to c Z)~(\sigma^{\mu\nu})$&   $2.3 \times 10^{-5}$&   $4.9 \times 10^{-5}$ &   $3.64 \times 10^{-6}$ &   $2.67 \times 10^{-5}$  &   $6.54 \times 10^{-7}$&     $2.61\times 10^{-5}$	\\
BR$(t \to u Z)~(\gamma^{\mu})$  &   $2.34 \times 10^{-5}$  &   $4.08\times 10^{-5}$ &   $4.28 \times 10^{-6}$ &   $2.47 \times 10^{-5}$&     $6.86\times 10^{-7}$&     $2.17 \times 10^{-5}$	\\
BR$(t \to c Z)~(\gamma^{\mu})$&   $3.13 \times 10^{-5}$&   $6.65 \times 10^{-5}$ &   $5.22 \times 10^{-6}$ &   $3.84 \times 10^{-5}$  &   $8.87 \times 10^{-7}$&     $3.54\times 10^{-5}$	\\	
\hline
\end{tabular}
\end{center}
\end{table*}

In Tab.~\ref{limit}, we list the exclusion limits at 95\% CL at the future HL-LHC with 3 ab$^{-1}$, at the  HE-LHC with 15 ab$^{-1}$ and at the FCC-hh with 30 ab$^{-1}$, respectively,  with two systematic error: $\delta=0\%$ and $\delta=10\%$.
From Tab.~\ref{limit}, we have the following observations to make,
 \begin{itemize}
\item
More stringent limits are obtained on the $tuZ$ coupling compared to the $tcZ$ coupling due to the larger cross section in the corresponding signal.
\item
For the $tuZ$ coupling, the sensitivities for the tensor couplings are smaller than those for the vector terms, being of the order of $10^{-6}$ at the 95\% CL by considering a $10\%$ systematic uncertainty.
\item
For both channels, the sensitivities are weaker than those without any systematic error. This means that those searches will be dominated by systematic uncertainties and
will not benefit further from the energy and luminosity upgrades.
\end{itemize}

Very recently, many phenomenological studies
available in literature have extensively investigated the top FCNC anomalous couplings at various future high energy colliders, including $e^+e^-$ and $e^-p$ machines: see, e.g. Refs.~\cite{Basso:2014apa,Aguilar-Saavedra:2017vka,Khanpour:2019qnw,Behera:2018ryv,Cakir:2018ruj,Khanpour:2014xla,AguilarSaavedra:2001ab,deBlas:2018mhx} for the most recent reviews. Besides, the expected
limits of the four-fermion coefficients at the LHeC and CEPC are obtained in Refs.~\cite{Shi:2019epw,Liu:2019wmi}.
It is then worth comparing the limits on ${\rm BR}(t\to qZ)$ obtained in this study with those obtained by other groups, which are summarised in Tab.~\ref{table-FCNC}. One can see that the limits on the BRs are expected to be of $\mathcal{O}(10^{-4}-10^{-6})$. Therefore, we expect our advocated signatures to provide  competitive complementary information to that from the above studies in detecting $tqZ$ ($q=u,c$) anomalous couplings at future hadronic colliders.

\begin{table}[htbp]
\renewcommand\arraystretch{0.9}
\begin{center}
 \caption{\label{table-FCNC}
Projected 95\% CL limits on ${\rm BR}(t\to qZ)$ ($q=u,c$) from different channels at varoius future colliders.}
\vspace{0.2cm}
\scalebox{0.95}{
\begin{tabular}{c| c| c}  \hline \hline
Channels      &Data Set &Limits  \\ \hline
\multirow{2}{*}{$tZ\to W(\to \ell \nu)bZ(\to \ell^{+}\ell^{-})$~\cite{Basso:2014apa}}&HL-LHC, 100 fb$^{-1}$  &${\rm BR}(t\to uZ)<1.6\times 10^{-4}$~($\sigma^{\mu\nu}$)\\
&@ 14 TeV&$ {\rm BR}(t\to cZ)< 1.0\times10^{-3}$~($\sigma^{\mu\nu}$)\\\cline{2-3} \hline
\multirow{4}{*}{Ultra-boosted $tZ$ production~\cite{Aguilar-Saavedra:2017vka}}&HL-LHC, 3 ab$^{-1}$  &${\rm BR}(t\to uZ)<4.1\times 10^{-5}$~($\sigma^{\mu\nu}$)\\
&@ 14 TeV&$ {\rm BR}(t\to cZ)< 1.6\times10^{-3}$~($\sigma^{\mu\nu}$)\\ \cline{2-3}
&FCC-hh, 10 ab$^{-1}$  &${\rm BR}(t\to uZ)< 2.7\times 10^{-6}$~($\sigma^{\mu\nu}$)\\
& @ 100 TeV&$ {\rm BR}(t\to cZ)< 5.0\times10^{-5}$~($\sigma^{\mu\nu}$)\\ \hline
\multirow{8}{*}{$pp\to  tt\bar{t}(\bar{t}\bar{t}t)$~\cite{Khanpour:2019qnw}} &\multirow{2}{*}{HE-LHC, 15 ab$^{-1}$}  &${\rm BR}(t\to uZ)< 2.4\times10^{-4}$~($\sigma^{\mu\nu}$)\\
&\multirow{2}{*}{@ 27 TeV}&${\rm BR}(t\to cZ)< 1.56\times 10^{-3}$~($\sigma^{\mu\nu}$)\\\cline{3-3}
&&${\rm BR}(t\to uZ)< 8.36\times10^{-4}$~($\gamma^{\mu}$)\\
&&${\rm BR}(t\to cZ)< 4.19\times 10^{-3}$~($\gamma^{\mu}$)\\ \cline{2-3}
&\multirow{2}{*}{FCC-hh, 10 ab$^{-1}$}  &${\rm BR}(t\to uZ)< 8.65\times10^{-5}$~($\sigma^{\mu\nu}$)~ \\
&\multirow{2}{*}{@ 100 TeV}&${\rm BR}(t\to cZ)< 2.33\times 10^{-4}$~($\sigma^{\mu\nu}$)\\ \cline{3-3}
&&${\rm BR}(t\to uZ)< 2.76\times10^{-4}$~($\gamma^{\mu}$)\\
&&${\rm BR}(t\to cZ)< 6.52\times 10^{-4}$~($\gamma^{\mu}$)\\
\hline
\multirow{4}{*}{$e^{-}p\to e^{-}t$~\cite{Behera:2018ryv} }&\multirow{2}{*}{ LHeC, 2 ab$^{-1}$}  &${\rm BR}(t\to uZ)<4\times 10^{-5}$~($\sigma^{\mu\nu}$)\\
&\multirow{2}{*}{ @ 60 GeV$ \oplus$ 7 TeV}&$ {\rm BR}(t\to cZ)< 6.8\times10^{-4}$~($\sigma^{\mu\nu}$)\\\cline{3-3}
&&${\rm BR}(t\to uZ)<9\times 10^{-5}$~($\gamma^{\mu}$)\\
&&$ {\rm BR}(t\to cZ)< 9.5\times10^{-4}$~($\gamma^{\mu}$)\\
\hline
\multirow{2}{*}{$e^{-}p\to e^{-}Wq$ + X~\cite{Cakir:2018ruj} }& LHeC, 3 ab$^{-1}$, $2\sigma$  &${\rm BR}(t\to qZ)<3.3\times 10^{-5}$~($\sigma^{\mu\nu}$)\\ \cline{2-3}
&FCC-he, 3 ab$^{-1}$, $2\sigma$  &${\rm BR}(t\to qZ)<4.5\times 10^{-6}$~($\sigma^{\mu\nu}$)\\ \hline
\multirow{2}{*}{$e^{+}e^{-}\to tq$~\cite{Khanpour:2014xla} }&FCC-ee, 300 fb$^{-1}$&${\rm BR}(t\to qZ)< 3.12\times 10^{-5}$~($\sigma^{\mu\nu}$)\\
& @ 350 GeV &${\rm BR}(t\to qZ)< 1.22\times 10^{-4}$~($\gamma^{\mu}$)\\ \hline
\multirow{2}{*}{$e^{+}e^{-}\to tq$~\cite{AguilarSaavedra:2001ab} }&ILC, 300 fb$^{-1}$&${\rm BR}(t\to qZ)< 1.9\times 10^{-3}$~($\sigma^{\mu\nu}$)\\
& @ 500 GeV &${\rm BR}(t\to qZ)< 1.8\times 10^{-3}$~($\gamma^{\mu}$)\\ \hline

\hline
 \end{tabular}}
 \end{center}
\end{table}

\section{Conclusions}
In this work, we have studied   FCNC $tZq$ anomalous couplings ($q=u,c$) at the future HL-LHC, HE-LHC and FCC-hh by performing a full simulation via two processes  yielding trilepton  signals: top quark pair production $pp\to t\bar{t}$ with $t\to qZ$ and the associated $tZ$ production process $pp\to tZ$.
We have performed a full simulation for the signals and the relevant SM backgrounds based on two separate cut selections
and obtained 95\% CL limits on ${\rm BR}(t\to qZ)$~$(q=u,c)$, by exploiting trilepton final states obtained via  the decay modes $t\to b W^{+}\to b\ell^{+}\nu_{\ell}$ and $Z\to \ell^{+}\ell^{-}$.
Altogether, these limits are nearly one or two orders of magnitude better than the current  experimental results obtained from LHC runs at 13 TeV. We therefore  expect that the signatures studied here will provide competitive complementary information for detecting such FCNC $tqZ$ anomalous couplings at future hadronic colliders at CERN.

\section*{Acknowledgments}
The work of Y.-B.L. is supported by the Foundation of the Henan Institute of Science and Technology (Grant no. 2016ZD01). S.M. is supported in part by the NExT Institute and the STFC CG Grant No. ST/L000296/1.

\end{document}